\input epsf
%
%
%
\def\unredoffs{} 

%
%
%
%
\newbox\leftpage \newdimen\fullhsize \newdimen\hstitle \newdimen\hsbody
\tolerance=1000\hfuzz=2pt
\catcode`\@=11 
%
\magnification=1200\unredoffs\baselineskip=16pt plus 2pt minus 1pt
\hsbody=\hsize \hstitle=\hsize 
%
%
%
\newcount\yearltd\yearltd=\year\advance\yearltd by -1900

%
%

\def\draftmode{\message{ DRAFTMODE }\def\draftdate{{\rm preliminary draft:
\number\month/\number\day/\number\yearltd\ \ \hourmin}}%
\headline={\hfil\draftdate}\writelabels\baselineskip=20pt plus 2pt minus 2pt
 {\count255=\time\divide\count255 by 60 \xdef\hourmin{\number\count255}
  \multiply\count255 by-60\advance\count255 by\time
  \xdef\hourmin{\hourmin:\ifnum\count255<10 0\fi\the\count255}}}
\def\nolabels{\def\wrlabeL##1{}\def\eqlabeL##1{}\def\reflabeL##1{}}
\def\writelabels{\def\wrlabeL##1{\leavevmode\vadjust{\rlap{\smash%
{\line{{\escapechar=` \hfill\rlap{\sevenrm\hskip.03in\string##1}}}}}}}%
\def\eqlabeL##1{{\escapechar-1\rlap{\sevenrm\hskip.05in\string##1}}}%
\def\reflabeL##1{\noexpand\llap{\noexpand\sevenrm\string\string\string##1}}}
\nolabels
%
\global\newcount\secno \global\secno=0
\global\newcount\meqno \global\meqno=1
\def\newsec#1{\global\advance\secno by1\message{(\the\secno. #1)}
\global\subsecno=0\eqnres@t\noindent{\bf\the\secno. #1}
\writetoca{{\secsym} {#1}}\par\nobreak\medskip\nobreak}
\def\eqnres@t{\xdef\secsym{\the\secno.}\global\meqno=1\bigbreak\bigskip}
\def\sequentialequations{\def\eqnres@t{\bigbreak}}\xdef\secsym{}
\global\newcount\subsecno \global\subsecno=0
\def\subsec#1{\global\advance\subsecno by1\message{(\secsym\the\subsecno. #1)}
\ifnum\lastpenalty>9000\else\bigbreak\fi
\noindent{\it\secsym\the\subsecno. #1}\writetoca{\string\quad
{\secsym\the\subsecno.} {#1}}\par\nobreak\medskip\nobreak}
\def\appendix#1#2{\global\meqno=1\global\subsecno=0\xdef\secsym{\hbox{#1.}}
\bigbreak\bigskip\noindent{\bf Appendix #1. #2}\message{(#1. #2)}
\writetoca{Appendix {#1.} {#2}}\par\nobreak\medskip\nobreak}
%
%
\def\eqnn#1{\xdef #1{(\secsym\the\meqno)}\writedef{#1\leftbracket#1}%
\global\advance\meqno by1\wrlabeL#1}
\def\eqna#1{\xdef #1##1{\hbox{$(\secsym\the\meqno##1)$}}
\writedef{#1\numbersign1\leftbracket#1{\numbersign1}}%
\global\advance\meqno by1\wrlabeL{#1$\{\}$}}
\def\eqn#1#2{\xdef #1{(\secsym\the\meqno)}\writedef{#1\leftbracket#1}%
\global\advance\meqno by1$$#2\eqno#1\eqlabeL#1$$}
%
\newskip\footskip\footskip14pt plus 1pt minus 1pt 
\def\footnotefont{\ninepoint}\def\f@t#1{\footnotefont #1\@foot}
\def\f@@t{\baselineskip\footskip\bgroup\footnotefont\aftergroup\@foot\let\next}
\setbox\strutbox=\hbox{\vrule height9.5pt depth4.5pt width0pt}
\global\newcount\ftno \global\ftno=0
\def\foot{\global\advance\ftno by1\footnote{$^{\the\ftno}$}}
%
\newwrite\ftfile
\def\footend{\def\foot{\global\advance\ftno by1\chardef\wfile=\ftfile
$^{\the\ftno}$\ifnum\ftno=1\immediate\openout\ftfile=foots.tmp\fi%
\immediate\write\ftfile{\noexpand\smallskip%
\noexpand\item{f\the\ftno:\ }\pctsign}\findarg}%
\def\footatend{\vfill\eject\immediate\closeout\ftfile{\parindent=20pt
\centerline{\bf Footnotes}\nobreak\bigskip\input foots.tmp }}}
\def\footatend{}
%
%
\global\newcount\refno \global\refno=1
\newwrite\rfile
\def\ref{[\the\refno]\nref}
\def\nref#1{\xdef#1{[\the\refno]}\writedef{#1\leftbracket#1}%
\ifnum\refno=1\immediate\openout\rfile=refs.tmp\fi
\global\advance\refno by1\chardef\wfile=\rfile\immediate
\write\rfile{\noexpand\item{#1\ }\reflabeL{#1\hskip.31in}\pctsign}\findarg}
\def\findarg#1#{\begingroup\obeylines\newlinechar=`\^^M\pass@rg}
{\obeylines\gdef\pass@rg#1{\writ@line\relax #1^^M\hbox{}^^M}%
\gdef\writ@line#1^^M{\expandafter\toks0\expandafter{\striprel@x #1}%
\edef\next{\the\toks0}\ifx\next\em@rk\let\next=\endgroup\else\ifx\next\empty%
\else\immediate\write\wfile{\the\toks0}\fi\let\next=\writ@line\fi\next\relax}}
\def\striprel@x#1{} \def\em@rk{\hbox{}}
\def\lref{\begingroup\obeylines\lr@f}
\def\lr@f#1#2{\gdef#1{\ref#1{#2}}\endgroup\unskip}
\def\semi{;\hfil\break}
\def\addref#1{\immediate\write\rfile{\noexpand\item{}#1}} 
\def\startrefs#1{\immediate\openout\rfile=refs.tmp\refno=#1}
\def\xref{\expandafter\xr@f}\def\xr@f[#1]{#1}
\def\refs#1{\count255=1[\r@fs #1{\hbox{}}]}
\def\r@fs#1{\ifx\und@fined#1\message{reflabel \string#1 is undefined.}%
\nref#1{need to supply reference \string#1.}\fi%
\vphantom{\hphantom{#1}}\edef\next{#1}\ifx\next\em@rk\def\next{}%
\else\ifx\next#1\ifodd\count255\relax\xref#1\count255=0\fi%
\else#1\count255=1\fi\let\next=\r@fs\fi\next}
%

%
\newwrite\ffile\global\newcount\figno \global\figno=1
\def\fig{fig.~\the\figno\nfig}
\def\nfig#1{\xdef#1{fig.~\the\figno}%
\writedef{#1\leftbracket fig.\noexpand~\the\figno}%
\ifnum\figno=1\immediate\openout\ffile=figs.tmp\fi\chardef\wfile=\ffile%
\immediate\write\ffile{\noexpand\medskip\noexpand\item{Fig.\ \the\figno. }
\reflabeL{#1\hskip.55in}\pctsign}\global\advance\figno by1\findarg}
\def\xfig{\expandafter\xf@g}\def\xf@g fig.\penalty\@M\ {}
\def\figs#1{figs.~\f@gs #1{\hbox{}}}
\def\f@gs#1{\edef\next{#1}\ifx\next\em@rk\def\next{}\else
\ifx\next#1\xfig #1\else#1\fi\let\next=\f@gs\fi\next}
\newwrite\lfile
{\escapechar-1\xdef\pctsign{\string\%}\xdef\leftbracket{\string\{}
\xdef\rightbracket{\string\}}\xdef\numbersign{\string\#}}

\def\writestop{\def\writestoppt{\immediate\write\lfile{\string\pageno%
\the\pageno\string\startrefs\leftbracket\the\refno\rightbracket%
\string\def\string\secsym\leftbracket\secsym\rightbracket%
\string\secno\the\secno\string\meqno\the\meqno}\immediate\closeout\lfile}}
\def\writestoppt{}\def\writedef#1{}
\def\seclab#1{\xdef #1{\the\secno}\writedef{#1\leftbracket#1}\wrlabeL{#1=#1}}
\def\subseclab#1{\xdef #1{\secsym\the\subsecno}%
\writedef{#1\leftbracket#1}\wrlabeL{#1=#1}}
\newwrite\tfile \def\writetoca#1{}
\def\leaderfill{\leaders\hbox to 1em{\hss.\hss}\hfill}
\def\writetoc{\immediate\openout\tfile=toc.tmp
   \def\writetoca##1{{\edef\next{\write\tfile{\noindent ##1
   \string\leaderfill {\noexpand\number\pageno} \par}}\next}}}

%
\catcode`\@=12 
%
\edef\tfontsize{\ifx\answ\bigans scaled\magstep3\else scaled\magstep4\fi}
\font\titlerm=cmr10 \tfontsize \font\titlerms=cmr7 \tfontsize
\font\titlermss=cmr5 \tfontsize \font\titlei=cmmi10 \tfontsize
\font\titleis=cmmi7 \tfontsize \font\titleiss=cmmi5 \tfontsize
\font\titlesy=cmsy10 \tfontsize \font\titlesys=cmsy7 \tfontsize
\font\titlesyss=cmsy5 \tfontsize \font\titleit=cmti10 \tfontsize
\skewchar\titlei='177 \skewchar\titleis='177 \skewchar\titleiss='177
\skewchar\titlesy='60 \skewchar\titlesys='60 \skewchar\titlesyss='60
\def\titlefont{\def\rm{\fam0\titlerm}
\textfont0=\titlerm \scriptfont0=\titlerms \scriptscriptfont0=\titlermss
\textfont1=\titlei \scriptfont1=\titleis \scriptscriptfont1=\titleiss
\textfont2=\titlesy \scriptfont2=\titlesys \scriptscriptfont2=\titlesyss
\textfont\itfam=\titleit \def\it{\fam\itfam\titleit}\rm}
 \ifx\answ\bigans\else scaled\magstep1\fi
\ifx\answ\bigans\else

 \font\absi=cmmi10 scaled\magstep1
\font\absis=cmmi7 scaled\magstep1 \font\absiss=cmmi5 scaled\magstep1
\font\abssy=cmsy10 scaled\magstep1 \font\abssys=cmsy7 scaled\magstep1
\font\abssyss=cmsy5 scaled\magstep1 
\skewchar\absi='177 \skewchar\absis='177 \skewchar\absiss='177
\skewchar\abssy='60 \skewchar\abssys='60 \skewchar\abssyss='60
\fi
\font\ninerm=cmr9 \font\sixrm=cmr6 \font\ninei=cmmi9 \font\sixi=cmmi6
\font\ninesy=cmsy9 \font\sixsy=cmsy6 \font\ninebf=cmbx9
\font\nineit=cmti9 \font\ninesl=cmsl9 \skewchar\ninei='177
\skewchar\sixi='177 \skewchar\ninesy='60 \skewchar\sixsy='60
\def\ninepoint{\def\rm{\fam0\ninerm}
\textfont0=\ninerm \scriptfont0=\sixrm \scriptscriptfont0=\fiverm
\textfont1=\ninei \scriptfont1=\sixi \scriptscriptfont1=\fivei
\textfont2=\ninesy \scriptfont2=\sixsy \scriptscriptfont2=\fivesy
\textfont\itfam=\ninei \def\it{\fam\itfam\nineit}\def\sl{\fam\slfam\ninesl}%
\textfont\bffam=\ninebf \def\bf{\fam\bffam\ninebf}\rm}
%
\def\noblackbox{\overfullrule=0pt}
\hyphenation{anom-aly anom-alies coun-ter-term coun-ter-terms}
\def\inv{^{\raise.15ex\hbox{${\scriptscriptstyle -}$}\kern-.05em 1}}

\def\Dsl{\,\raise.15ex\hbox{/}\mkern-13.5mu D} 
\def\dsl{\raise.15ex\hbox{/}\kern-.57em\partial}

\def\lspace{\ifx\answ\bigans{}\else\qquad\fi}
\def\lbspace{\ifx\answ\bigans{}\else\hskip-.2in\fi} 
\def\boxeqn#1{\vcenter{\vbox{\hrule\hbox{\vrule\kern3pt\vbox{\kern3pt
        \hbox{${\displaystyle #1}$}\kern3pt}\kern3pt\vrule}\hrule}}}
\def\mbox#1#2{\vcenter{\hrule \hbox{\vrule height#2in
                \kern#1in \vrule} \hrule}}  
%
   
 \def\CH{{\cal H}}

\def\darr#1{\raise1.5ex\hbox{$\leftrightarrow$}\mkern-16.5mu #1}

\def\half{{\textstyle{1\over2}}} 
\def\roughly#1{\raise.3ex\hbox{$#1$\kern-.75em\lower1ex\hbox{$\sim$}}}
\hyphenation{Mar-ti-nel-li}
\def\hp{{\hat p}}
\def\hq{{\hat q}}

\def\1{\;1\!\!\!\! 1\;}

\def\etal{{\it et al.}}

\def\frac#1#2{{{#1}\over {#2}}}
\def\half{\hbox{${1\over 2}$}}\def\third{\hbox{${1\over 3}$}}
\def\quarter{\hbox{${1\over 4}$}}
\def\eigth{\hbox{${1\over 8}$}}
\def\smallfrac#1#2{\hbox{${{#1}\over {#2}}$}}

\def\MS{\hbox{$\overline{\rm MS}$}}
\def\ms{\hbox{$\overline{\scriptstyle\rm MS}$}}
\def\MSS{\hbox{$\overline{\rm MS}^*$}}
\def\mss{\hbox{$\overline{\scriptstyle\rm MS}^*$}}

\def\Q0{\hbox{$\rm Q_0$}}
\def\q0{\hbox{$\scriptstyle\rm Q_0$}}

\catcode`@=11 
\def\slash#1{\mathord{\mathpalette\c@ncel#1}}
 \def\c@ncel#1#2{\ooalign{$\hfil#1\mkern1mu/\hfil$\crcr$#1#2$}}
\def\lsim{\mathrel{\mathpalette\@versim<}}
\def\gsim{\mathrel{\mathpalette\@versim>}}
 \def\@versim#1#2{\lower0.2ex\vbox{\baselineskip\z@skip\lineskip\z@skip
       \lineskiplimit\z@\ialign{$\m@th#1\hfil##$\crcr#2\crcr\sim\crcr}}}
\catcode`@=12 

\def\PR{{\it Phys.~Rev.~}}

\def\NP{{\it Nucl.~Phys.~}}

\def\PL{{\it Phys.~Lett.~}}

\def\SJNP{{\it Sov.~Jour.~Nucl.~Phys.~}}
\def\SPJETP{{\it Sov.~Phys.~J.E.T.P.~}}

\def\JHEP{{\it Jour.~High~Energy~Phys.~}}
\def\vol#1{{\bf #1}}\def\vyp#1#2#3{\vol{#1} (#2) #3}

\def\as{\alpha_s}
\def\ahat{\hat\as}
\def\ahati{{\hat\as^{-1}}}

\def\ash{\widehat\alpha_s}

\noblackbox
\pageno=0\nopagenumbers\tolerance=10000\hfuzz=5pt
\baselineskip 12pt
\line{\hfill Edinburgh 2007/08}
\line{\hfill IFUM-881-FT}
\vskip 12pt
\centerline{\titlefont BFKL at Next-to-Next-to-Leading Order}
\vskip 36pt\centerline{Simone Marzani$^{(a)}$, Richard D.~Ball$^{(a)}$, Pietro
  Falgari$^{(b)}$ and Stefano Forte$^{(c)}$}
\vskip 12pt
\centerline{\it ${}^{(a)}$School of Physics, University of Edinburgh}
\centerline{\it  Edinburgh EH9 3JZ, Scotland}
\vskip 6pt
\centerline {\it ${}^{(b)}$Institut f\"ur Theoretische Physik E, RWTH Aachen,}
\centerline{\it D-52056 Aachen, Germany}
\vskip 6pt
\centerline {\it ${}^{(c)}$Dipartimento di  Fisica, Universit\`a di
Milano and}
\centerline{\it INFN, Sezione di Milano, Via Celoria 16, I-20133 Milan, Italy}
\vskip 54pt
\centerline{\bf Abstract}
{\narrower\baselineskip 10pt
\medskip\noindent
We determine an approximate expression for the $O(\as^3)$
contribution $\chi_2$ to the kernel of the BFKL equation, which 
includes all collinear and anticollinear singular contributions. 
This is derived using recent 
results on the relation between the GLAP and
BFKL kernels (including running-coupling effects to all orders)
and on small-$x$ factorization schemes. We present the result in
various schemes, relevant both for applications to the BFKL
equation and to small-$x$ evolution of parton distributions. }
\vfill \line{April 2007\hfill} \eject
\footline={\hss\tenrm\folio\hss}

\lref\bch{E.~Eriksen, {\it J. Math. Phys.}\vyp{9}{1969}{790}}
\lref\brus  {R.~D.~Ball and S.~Forte,
  {\tt hep-ph/9805315.}
}
\lref\fal{P.~Falgari, {\it Laurea Thesis}, Milan University, April 2005}
\lref\mar{S.~Marzani, {\it Laurea Thesis}, Milan University, April 2005}
\lref\marfal{R.~D.~Ball, P.~Falgari, S.~Forte and S.~Marzani, {\it in
  preparation} }
\lref\glap{
V.N.~Gribov and L.N.~Lipatov,
\SJNP\vyp{15}{1972}{438}\semi  
L.N.~Lipatov, \SJNP\vyp{20}{1975}{95}\semi    
G.~Altarelli and G.~Parisi,
\NP\vyp{B126}{1977}{298}\semi  
see also
Y.L.~Dokshitzer,
{\it Sov.~Phys.~JETP~}\vyp{46}{1977}{691}.} 
\lref\cfp{G.~Curci, W.~Furma\'nski and R.~Petronzio,
\NP\vyp{B175}{1980}{27}} 
\lref\nlo{E.G.~Floratos, C.~Kounnas and R.~Lacaze,
\NP\vyp{B192}{1981}{417}.} 
\lref\nnlo{S.A.~Larin, T.~van~Ritbergen, J.A.M.~Vermaseren,
\NP\vyp{B427}{1994}{41}\semi  
S.A.~Larin \etal, \NP\vyp{B492}{1997}{338}.} 
\lref\bfkl{L.N.~Lipatov,
\SJNP\vyp{23}{1976}{338}\semi 
 V.S.~Fadin, E.A.~Kuraev and L.N.~Lipatov,
\PL\vyp{60B}{1975}{50}; 
 {\it Sov. Phys. JETP~}\vyp{44}{1976}{443}; 
\vyp{45}{1977}{199}\semi 
 Y.Y.~Balitski and L.N.Lipatov,
\SJNP\vyp{28}{1978}{822}.} 
\lref\CH{
S.~Catani and F.~Hautmann,
\PL\vyp{B315}{1993}{157}; 
\NP\vyp{B427}{1994}{475}.} 
\lref\fl{V.S.~Fadin and L.N.~Lipatov,
\PL\vyp{B429}{1998}{127}.  
}
\lref\fleta{
V.S.~Fadin et al, \PL\vyp{B359}{1995}{181}; 
\PL\vyp{B387}{1996}{593}; 
\NP\vyp{B406}{1993}{259}; 
\PR\vyp{D50}{1994}{5893}; 
\PL\vyp{B389}{1996}{737};  
\NP\vyp{B477}{1996}{767};  
\PL\vyp{B415}{1997}{97};  
\PL\vyp{B422}{1998}{287}.} 
\lref\cc{G.~Camici and M.~Ciafaloni,
\PL\vyp{B412}{1997}{396}; 
\PL\vyp{B430}{1998}{349}.} 
\lref\dd{V.~del~Duca, \PR\vyp{D54}{1996}{989};
\PR\vyp{D54}{1996}{4474}\semi 
V.~del~Duca and C.R.~Schmidt,
\PR\vyp{D57}{1998}{4069}\semi 
Z.~Bern, V.~del~Duca and C.R.~Schmidt,
\PL\vyp{B445}{1998}{168}.}
\lref\ross{
D.~A.~Ross,
Phys.\ Lett.\ B {\bf 431}, 161 (1998) 
}
\lref\jar{T.~Jaroszewicz,
\PL\vyp{B116}{1982}{291}.}
\lref\ktfact{  S.~Catani, M.~Ciafaloni and F.~Hautmann,
  Phys.\ Lett.\ B {\bf 242}, 97 (1990).
}
\lref\ktfaca{S.~Catani, F.~Fiorani and G.~Marchesini,
\PL\vyp{B336}{1990}{18}\semi 
S.~Catani et al.,
\NP\vyp{B361}{1991}{645}.}
\lref\summ{R.~D.~Ball and S.~Forte,
\PL\vyp{B351}{1995}{313}\semi  
R.K.~Ellis, F.~Hautmann and B.R.~Webber,
\PL\vyp{B348}{1995}{582}.}
\lref\afp{R.~D.~Ball and S.~Forte,
\PL\vyp{B405}{1997}{317}.}
\lref\DGPTWZ{A.~De~R\'ujula {\it et al.},
\PR\vyp{D10}{1974}{1649}.}
\lref\das{R.~D.~Ball and S.~Forte,
\PL\vyp{B335}{1994}{77}; 
\vyp{B336}{1994}{77}\semi 
{\it Acta~Phys.~Polon.~}\vyp{B26}{1995}{2097}.}
\lref\kis{
See {\it e.g.}  R.~K.~Ellis, W.~J.~Stirling and B.~R.~Webber,
``QCD and Collider Physics'' (C.U.P., Cambridge 1996).}
\lref\hone{H1 Collab., {\it Eur.\ Phys.\ J.} C {\bf 21} (2001)
33.}
\lref\zeus{ZEUS Collab., {\it Eur.\ Phys.\ J.}
 C {\bf 21} (2001) 443.} 
\lref\mom{R.~D.~Ball and S.~Forte, {\it Phys. Lett.} {\bf
B359}, 362 (1995).}
\lref\bfklfits{R.~D.~Ball and S.~Forte,
{\tt hep-ph/9607291}\semi 
I.~Bojak and M.~Ernst, \PL\vyp{B397}{1997}{296};
\NP\vyp{B508}{1997}{731}\semi
J.~Bl\"umlein  and A.~Vogt,
\PR\vyp{D58}{1998}{014020}.}
\lref\flph{R.~D.~Ball  and S.~Forte,
{\tt hep-ph/9805315}\semi 
J. Bl\"umlein et al.,
{\tt hep-ph/9806368}.}
\lref\salam{G.~Salam, \JHEP\vyp{9807}{1998}{19}.}
\lref\sxap{R.~D.~Ball and S.~Forte,
\PL\vyp{B465}{1999}{271}.}
\lref\sxres{G. Altarelli, R.~D. Ball and S. Forte,
\NP{\bf B575}, 313 (2000);  
see also {\tt hep-ph/0001157}.
}
\lref\sxphen{G. Altarelli, R.~D.~Ball and S. Forte,
\NP\vyp{B599}{2001}{383};  
see also {\tt hep-ph/0104246}.}  
\lref\tw{ C.~D.~White and R.~S.~Thorne,
  \PR\vyp{D75}{2007}{034005}.
}
\lref\ciaf{M.~Ciafaloni and D.~Colferai,
\PL\vyp{B452}{1999}{372}. 
}
\lref\ciafresa{
 M.~Ciafaloni, D.~Colferai and G.~P.~Salam,
  Phys.\ Rev.\ D {\bf 60} (1999) 114036.
}
\lref\ciafresb{M.~Ciafaloni, D.~Colferai, G.~P.~Salam and A.~M.~Stasto,
  Phys.\ Rev.\ D {\bf 66} (2002) 054014;
  Phys.\ Lett.\ B {\bf 576} (2003) 143.
}
\lref\ciafres{ M.~Ciafaloni, D.~Colferai and G.~P.~Salam,
  Phys.\ Rev.\ D {\bf 60} (1999) 114036;
M.~Ciafaloni, D.~Colferai, G.~P.~Salam and A.~M.~Stasto,
  Phys.\ Rev.\ D {\bf 68} (2003) 114003.
}
\lref\heralhcres{G.~Altarelli {\it et al.}, ``Resummation'', in M.~Dittmar {\it et al.},
  {\tt hep-ph/0511119.}}
\lref\heralhc{ M.~Dittmar {\it et al.},
  {\tt hep-ph/0511119.}
}
\lref\ciafdip{M.~Ciafaloni, D.~Colferai, G.~P.~Salam and A.~M.~Stasto,
  Phys.\ Lett.\ B {\bf 587} (2004) 87; see also
 G.~P.~Salam,
{\tt hep-ph/0501097}.

}
\lref\sxsym{G.~Altarelli, R.~D.~Ball and S.~Forte,
 {\tt hep-ph/0512237}.
}
\lref\Liprun{L.N.~Lipatov,
\SPJETP\vyp{63}{1986}{5}.}
\lref\ColKwie{
J.~C.~Collins and J.~Kwiecinski, \NP\vyp{B316}{1989}{307}.}
\lref\CiaMue{
M.~Ciafaloni, M.~Taiuti and A.~H.~Mueller,
{\tt hep-ph/0107009}.
}
\lref\ciafac{
M.~Ciafaloni, D.~Colferai and G.~P.~Salam,
JHEP {\bf 0007}  (2000) 054
}
\lref\ciafrun{G.~Camici and M.~Ciafaloni,
\NP\vyp{B496}{1997}{305}.}
\lref\Haak{L.~P.~A.~Haakman, O.~V.~Kancheli and
J.~H.~Koch \NP\vyp{B518}{1998}{275}.} 
\lref\Bartels{N. Armesto, J. Bartels and M.~A.~Braun,
\PL\vyp{B442}{1998}{459}.} 
\lref\Thorne{R.~S.~Thorne,
\PL\vyp{B474}{2000}{372}; {\it Phys.\ Rev.} {\bf D64} (2001) 074005.
}
\lref\anders{
J.~R.~Andersen and A.~Sabio Vera,
{\tt arXiv:hep-ph/0305236.}
}
\lref\mf{
S.~Forte and R.~D.~Ball,
{\tt hep-ph/0109235.}
}
\lref\sxrun{
G.~Altarelli, R.~D.~Ball and S.~Forte,
Nucl.\ Phys.\ B {\bf 621} (2002)  359.
}
\lref\ciafqz{
  M.~Ciafaloni, 
  {\tt hep-th/9510025};
  Phys.\ Lett.\ B {\bf 356}, 74 (1995).
}
\lref\ciafmsb{S.~Catani, M.~Ciafaloni and F.~Hautmann,
  Phys.\ Lett.\ B {\bf 307}, 147 (1993).
}
\lref\ciafrc{M.~Ciafaloni, M.~Taiuti and A.~H.~Mueller,
{\it Nucl.\ Phys.}  {\bf B616} (2001) 349\semi 
M.~Ciafaloni et al., {\it Phys. Rev.} {\bf D66} (2002)
054014 
}
\lref\runph{G.~Altarelli, R.~D.~Ball and S.~Forte,
Nucl.\ Phys.\ B {\bf 674} (2003) 459;
see also   
 {\tt hep-ph/0310016.}
}
\lref\symres{G.~Altarelli, R.~D.~Ball and S.~Forte,
  Nucl.\ Phys.\ Proc.\ Suppl.\  {\bf 135} (2004) 163
}
\lref\nnlo{ S.~Moch, J.~A.~M.~Vermaseren and A.~Vogt,
  Nucl.\ Phys.\ B {\bf 691} (2004) 129\semi
 A.~Vogt, S.~Moch and J.~A.~M.~Vermaseren,
  Nucl.\ Phys.\ B {\bf 688} (2004) 101.
}
\lref\sxsym{
  G.~Altarelli, R.~D.~Ball and S.~Forte,
  Nucl.\ Phys.\ B {\bf 742}, 1 (2006); see also
  S.~Forte, G.~Altarelli and R.~D.~Ball,
  {\tt hep-ph/0606323}.
}

\lref\mrst{See e.g. R.~S.~Thorne, A.~D.~Martin, R.~G.~Roberts and W.~J.~Stirling,
  {\tt hep-ph/0507015.}
}
\lref\ciafscheme{M.~Ciafaloni and D.~Colferai,
  JHEP {\bf 0509}, 069 (2005).
}
\lref\ciafschemph{
  M.~Ciafaloni, D.~Colferai, G.~P.~Salam and A.~M.~Stasto,
  Phys.\ Lett.\ B {\bf 635}, 320 (2006).
}
\lref\bfcomm{
  R.~D.~Ball and S.~Forte,
  Nucl.\ Phys.\ B {\bf 742}, 158 (2006).
}

\lref\andersen{J.~R.~Andersen and A.~Sabio Vera,
  Nucl.\ Phys.\  B {\bf 679} (2004) 345.
}
\lref\peps{ R.~K.~Ellis, D.~A.~Ross and A.~E.~Terrano,
  Nucl.\ Phys.\  B {\bf 178} (1981) 421.
}
\lref\dynlo{G.~Altarelli, R.~K.~Ellis and G.~Martinelli,
  Nucl.\ Phys.\  B {\bf 143} (1978) 521
  [Erratum-ibid.\  B {\bf 146} (1978) 544];  
  Nucl.\ Phys.\  B {\bf 157} (1979) 461.
}
\lref\hnlo{  S.~Dawson,
  Nucl.\ Phys.\  B {\bf 359} (1991) 283;
 A.~Djouadi, M.~Spira and P.~M.~Zerwas,
  Phys.\ Lett.\  B {\bf 264} (1991) 440.
}
\lref\kos{ D.~A.~Kosower and P.~Uwer,
  Nucl.\ Phys.\  B {\bf 563} (1999) 477.
}
\lref\lund{ J.~R.~Andersen {\it et al.}  [Small x Collaboration],
  Eur.\ Phys.\ J.\  C {\bf 48} (2006) 53.
}
\lref\enterrev{D.~d'Enterria,
{\tt hep-ex/0610061}.
}
\lref\salamrev{See e.g.
  G.~P.~Salam,
  {\tt hep-ph/0607153}.}
\lref\vogtapp{W.~L.~van Neerven and A.~Vogt,
  Nucl.\ Phys.\ B {\bf 568}, 263 (2000);
  Nucl.\ Phys.\ B {\bf 588}, 345 (2000);
  Phys.\ Lett.\ B {\bf 490}, 111 (2000).
}
\lref\regone{ V.~Del Duca and E.~W.~N.~Glover,
  JHEP {\bf 0110}, 035 (2001).}
\lref\regtwo{ 
  A.~V.~Bogdan, V.~Del Duca, V.~S.~Fadin and E.~W.~N.~Glover,
  JHEP {\bf 0203} (2002) 032.}
\lref\largexresa{
  S.~Albino and R.~D.~Ball,
  Phys.\ Lett.\  B {\bf 513} (2001) 93.}
\lref\largexres{
  P.~Bolzoni, S.~Forte and G.~Ridolfi,
  Nucl.\ Phys.\ B {\bf 731}, 85 (2005).}

\newsec{The BFKL equation beyond next-to-leading order}
\noindent Higher order calculations in perturbative QCD, both at
fixed order and at the resummed level, are playing an increasingly
important role in precision collider physics ~\salamrev.
Fixed-order and resummed results pose important constraints on
each other. On the one hand, fixed-order computations determine an
infinite number of coefficients of resummed expressions: in the
soft (e.g. large-$x$) limit ~\refs{\largexresa,\largexres}, 
the resummed hard cross-section 
can be determined to any desired
logarithmic order by a finite fixed-order
computation of suitably high accuracy, 
while at high energy (e.g. small $x$) there is a duality~\afp\ such that a
finite-order computation of the GLAP kernel determines an
all-order resummation of the BFKL kernel and conversely. On the
other hand, available resummed results determine partly higher
fixed-order expressions and allow an approximate reconstruction of
their form: in fact,  approximate determinations of the
next-to-next-to-leading order GLAP splitting functions~\vogtapp,
which played a useful role phenomenological until the exact
expressions became available\nnlo, are significantly
constrained by knowledge of the large-$x$ and small-$x$ behaviour
from resummed results.

The interplay of finite-order and resummed results is especially
interesting for the high-energy (small-$x$) limit of hard
cross-sections, which behave as genuine two-scale processes. The
dependence on the hard scale (henceforth $Q^2$, for definiteness)
and the energy scale (actually, the dimensionless ratio, $x$ for
definiteness, that controls the energy dependence) are governed by
a pair of evolution equations, the GLAP and BFKL equations,
respectively, whose kernels are related by a duality relation
which was recently shown to hold to all orders at the
running--coupling level~\bfcomm. This duality relation determines
the resummed expansion of either kernel in terms of the
fixed-order expansion of the other. Hence, even when the symmetry
between the two scales is broken by the running of the coupling
(and/or by kinematics) knowledge of either of the two kernels at
fixed orders enables an all-order resummation of the other kernel.

These results have been mainly used to perform small-$x$
resummation of GLAP evolution (see refs.~\refs{\sxsym,\ciafres}
and refs. therein, and refs.~\refs{\sxphen,\tw} for phenomenological
applications), 
i.e. to learn about higher-order
contributions to the GLAP kernel. However, they can also be used
to determine higher-order contributions to the BFKL kernel:
indeed, they provided a nontrivial check on the next-to-leading
order determination of the BFKL kernel~\refs{\fl,\cc}. Their use
to determine corrections to the BFKL kernel beyond next-to-leading
order, which are hitherto unknown, has been hampered by two
difficulties. First, whereas some next-to-next-to leading order
duality relations have been worked out some time ago~\mf, they
didn't include the full next-to-leading order running of the
coupling, nor the information contained in  the then unknown NNLO
GLAP kernel. In fact, the inclusion of these contributions is quite
hard if the running--coupling duality relations are solved
by brute force, even using computer algebra as in Ref.~\mf.
Second, beyond leading order both fixed-order and resummed
results, or, equivalently, both the BFKL kernel and the GLAP
kernel depend on a choice of factorization scheme. Factorization
schemes at small-$x$ were determined explicitly in
refs.~\refs{\ciafmsb,\ciafqz} and their implications for the
GLAP-BFKL duality were worked out in refs.~\refs{\mom,\sxphen},
but only up to next-to-leading order.

Recent results solve both difficulties. In ref.~\bfcomm\ a general
method has been developed which allows an efficient determination
of duality relations  by purely algebraic techniques, with full
inclusion of the running of the coupling  to any desired order.
Thanks to the full computation of the GLAP splitting functions to
next-to-next-to leading order,~\nnlo\ a computation of all the
singular contributions to the next-to-next-to leading order BFKL
kernel $\chi_2$ is now possible. Furthemore in ref.~\ciafscheme\
small-$x$ scheme changes have been discussed to all orders. Even
though in ref.~\ciafscheme\ the scheme change required for the
BFKL-GLAP is determined explicitly only up to NLO, it turns out
that its determination up to NNLO is possible from available
results, as we shall discuss below, at least for the terms
which affect the singular contributions to the NNLO BFKL kernel.
The purpose of the present paper is to perform these computations,
and use them to construct an approximate form of the $\chi_2$
kernel.

A determination of the BFKL kernel to NNLO is of theoretical and
phenomenological interest for various reasons. It is well-known
that NLO corrections to the BFKL kernel are large and in fact
change completely its qualitative shape. The determination of the
NNLO contribution is thus motivated not only by the slow
convergence of the perturbative expansion of the BFKL kernel, but
also by the expectation that (because of the alternating sign of
the dominant contributions) the NNLO approximation has a minimum like
the LO, and thus 
better stability properties than the NLO. Also, it is unclear
whether a direct extraction of the NNLO BFKL kernel from the
high-energy behaviour of parton-parton scattering amplitudes
analogous to the NLO computation of Ref.~\fl\ is feasible, because
it is unclear whether beyond NLO some form of ``reggeization'' 
holds, i.e. whether the exchange of an effective colorless multigluon
state is universal \refs{\regone,\regtwo}. If this is not 
the case, a derivation of the NNLO BFKL kernel from
high-energy factorization may be the only viable option.

This paper will be organized as follows. In the next section we
will summarize the formalism of Ref.~\bfcomm\ for the algebraic
resolution of duality relations between the GLAP and BFKL kernel,
and describe specifically its application to the extraction of all
available information on the NNLO BFKL kernel from the known NNLO
GLAP result. Then in Section~3 we will turn to the issue of scheme
dependence at small $x$ at NNLO: after summarizing the general
results of ref.~\ciafscheme, we will describe its application to
NNLO duality. In Section~4 we will finally determine explicitly
the approximate NNLO $\chi_2$ kernel in various relevant
factorization schemes, discuss its features and estimate the
accuracy of our approximation. Technical results on higher-order
dualities and on the so-called $Q_0$ scheme at NNLO are collected
in the appendices.

\newsec{The GLAP--BFKL duality}
\noindent Duality is the statement that the solutions to the GLAP
and BFKL equations coincide up to higher-twist corrections
provided their kernels are suitably matched. Its consequence is
that the leading-twist part of each kernel is determined by the
other kernel. This result is straightforward to establish at fixed
coupling~\refs{\jar,\afp,\sxres}, but rather more subtle when the
coupling runs~\refs{\sxrun,\bfcomm}. Here we summarize the main
results, while referring to Ref.~\bfcomm\ for a more comprehensive
treatment.

We discuss evolution equations for a parton distribution $G(x,
Q^2)$, which can be thought of simply as the gluon density, or as an
eigenvector of a two-by-two evolution matrix in the singlet
sector. The kinematic variables $x$ and $Q^2$ can be thought of as
the standard DIS variables, or more generally the perturbative
scale and a dimensionless scale ratio such that $0\le x\le 1$. We
will not consider the dependence on other kinematic variables,
such as transverse momentum and rapidity, i.e. we will consider
standard parton distributions, for which ordinary collinear
factorization applies. We will therefore limit ourselves to
angular-averaged quantities at the leading-twist level. We express
the parton density as a function of the logs of the relevant
kinematic variables: \eqn\baskin {G=G(\xi,t);\qquad
\xi\equiv\log{1\over x},\quad t\equiv \log{Q^2\over \mu^2}, } and
define the Mellin transform with respect to either (or both) of
the kinematic variables:\eqnn\nmom\eqnn\mmom
$$ \eqalignno{G(N,t) &\equiv\int^{\infty}_{0}\! d\xi\, e^{-N\xi}
 G(\xi,t),  &\nmom\cr
 G(\xi,M)&\equiv\int^{\infty}_{-\infty}\! dt\, e^{-Mt}
 G(\xi,t). &\mmom\cr
}$$
Note that, by slight abuse of notation, we denote with the
same symbol the parton distributions $G(N,t)$, $G(\xi,M)$, and
$G(N,M)$, although they are of course different functions of the
respective arguments.

The GLAP equation and BFKL equations express respectively the $t$
or $\xi$ dependence of $G$. They take the form
\eqnn\tevol\eqnn\xevol
$$\eqalignno
{\frac{d}{dt}G(N,t)&=\gamma(\as(t),N) G(N,t),&\tevol \cr
\frac{d}{d\xi}G(\xi,M)&=\chi(\ahat,M) G(\xi,M),&\xevol\cr}$$ where
$\as(t)$ is the running coupling, which upon Mellin transformation
becomes the operator $\ahat$, obtained by replacing $t\to
-{\partial\over \partial M}$ in the expression for $\as(t)$. For
example, at the leading-log level, where $\as(t)=\as
/(1+\beta_0\as t) $, \eqn\ashdef{ \ash = \frac{\as}{1-\beta_0 \as
\smallfrac{\partial}{\partial M}}, } with $\as\equiv\as(0)$.

At {\it fixed} coupling it is easy to show that if the kernels
$\chi$ and $\gamma$ are related by \eqnn\dualdef\eqnn\dualinv
$$\eqalignno{
N&=\chi(\as,\gamma(\as,N)),&\dualdef\cr
M&=\gamma(\as,\chi(\as,M)), &\dualinv\cr}
$$
then the BFKL and GLAP equation admit the same solution. This
relation is straightforward to derive~\refs{\sxap,\sxres} from the
observation that the leading-twist behaviour of $G(N,M)$ is
determined by its pole in the $(M,N)$ plane, and that
eqs.~\dualdef\ and \dualinv\ express the position of this pole. This
duality maps the expansion of $\chi(\as,M)$ in powers of $\as$ at
fixed $M$ onto the expansion of $\gamma(\as,N)$ in powers of $\as$
at fixed $\as/N$, and the expansion of $\gamma(\as,N)$ in powers
of $\as$ at fixed $N$ onto the expansion of $\chi(\as,M)$ in
powers of $\as$ at fixed $\as/M$.

At the running--coupling level, duality relations can be derived
order by order by solving eqs.~\tevol\ and \xevol\ perturbatively and
matching the respective solutions~\mf. That this is possible
beyond next-to-leading order is highly nontrivial, and it can be
proven to all orders using operator methods~\bfcomm. Namely, one
shows that at the running--coupling level, the BFKL and GLAP
solutions coincide if the respective kernels, viewed as operators
in $(M,N)$ space, are the inverse of each other when acting on
physical states, i.e. such that if \eqn\polecg{
MG(N,M)=\gamma(\ahat,N)G(N,M),} then \eqn\polecc{
NG(N,M)=\chi(\ahat,M)G(N,M),} and conversely.
Note that the conditions eq.~\polecg\ and \polecc\ should be viewed as
conditions on the action of the operators $\gamma(\ahat, N)$ and
$\chi(\ahat,M)$ on physical states: specifically, they are not just the
Mellin transforms of eqs.~\tevol,\xevol, which depend on a boundary
condition, but rather, they express a property of the leading--twist
Green function of perturbative GLAP or BFKL evolution~\bfcomm.

The inversion is nontrivial because the operators involved do not
commute. Indeed, we can start with eq.~\polecg\ and construct
$\tilde\chi(\ahat,M)$ as a series in $M$ such that
\eqn\gamopd{\tilde\chi(\ahat,\gamma(\ahat,N))G(N,M)=NG(N,M).}
Because $\ahat$ and $N$ commute, $\tilde\chi(\ahat,M)$ coincides
with the inverse function of $\gamma$  i.e. the fixed-coupling (or
naive) dual eq.~\dualdef: it is the same power series in $M$.
However, because $\gamma(\ahat,N)$ and $M$ do {\it not} commute,
we cannot use eq.~\polecg\ to identify $\gamma(\ahat,N)$ with $M$
in eq.~\gamopd, and thus obtain the desired equation~\polecc.

The  problem can be formalized as follows: given an  operator
equation of the form 
\eqn\opeq{\hq G= \hp G,} 
and given a function
$f(\hat q)$, determine the function $g$ such that using eq.~\opeq\
one gets \eqn\opeqb{f(\hat q)G=g(\hat p)G.} It is easy to see that
when $\hat p$ and $\hat q$ do not commute, the functions $f$ and
$g$ do not coincide: in fact they can be determined in terms of
each other using the expansion of the Baker-Campbell-Hausdorff
formula~\bch\ to lowest nontrivial order: \eqn\eefopeq{
f(\hat{q})G=
\big\{f(\hat{p})-\smallfrac{1}{2}f''(\hat{p})\left[\hat{p},\hat{q}\right]
+\ldots\big\}G, } so that \eqn\gexp{ g(\hat{p})=
f(\hat{p})-\smallfrac{1}{2}f''(\hat{p})\left[\hat{p},\hat{q}\right]
+\ldots.} Expressions up to third order are derived in
ref.~\bfcomm, and higher-order expressions (up to fifth order) are
given in Appendix~A.

Hence, if we identify \eqn\opidinv{\hat p=M,\qquad\qquad\hat
q=\gamma(\ahat,N),} and then choose $f=\tilde\chi$,
we may use  eq.~\gexp\ (and its higher-order generalizations) to
determine $\tilde\chi(\ahat,\gamma(\ahat,N))$ on the l.h.s. of
eq.~\gamopd\ as a function of $M$, which we then identify with the
sought-for BFKL kernel $\chi(\ahat,M)$. The computation of
$\chi(\ahat,M)$ is thus reduced by eq.~\eefopeq\ to the
determination of commutators, which in our case can all be
obtained from the basic commutator
\eqn\ascomm{[\ahati,M]=-\beta_0+\as\beta_0\beta_1+\dots,} where
the QCD beta function is given by
\eqn\betafun{\beta(\as)=-\beta_0\as^2(1+\beta_1\as)+\dots .} Using
this commutator in eq.~\eefopeq, to lowest nontrivial order we get
\eqn\rcnlo{ N G(N,M) =  \left[\tilde\chi(\ahat,M) +
\frac{1}{2}\beta_{0}\ahat^2 {\partial\over\partial\ahat}\gamma(\ahat,N)\,
\tilde\chi''(\ahat,M)\right]G(N,M)+ O(\ahat^2),} where 
primes denote derivatives with respect to the second argument ($M$ or
$N$).

In order for the r.h.s. of eq.~\rcnlo\ to provide us with an
expression for $\chi(\ahat,M)$ we must eliminate the residual $N$
dependence in it. This can be done in two steps. First, we can
back-substitute the lower-order expressions for $N$ given by
eq.~\rcnlo\ in its higher-order terms: for example, eq.~\rcnlo\
tells us that we can just replace the leading order expression
$\tilde\chi(\ahat,M)$ for the $N$ dependence of the
next-to-leading order $O(\beta_0)$ term, up to $O(\ahat^2)$
corrections: \eqn\nloback{ N G(N,M) =  \left[\tilde\chi(\ahat,M) +
\frac{1}{2}\beta_{0}\ahat^2 {\partial\over\partial\ahat}
{\gamma}(\ahat,\tilde\chi(\ahat,M)) 
\,\tilde\chi''(\ahat,M)\right]G(N,M)+O(\ahat^2). }
 Of course, beyond next-to-leading order this back-substitution becomes
nontrivial, and it must be performed by keeping into account the
commutation properties of the operators involved.

The dependence of the result eq.~\nloback\ on
$\gamma(\ahat,\tilde\chi(\ahat,M ))$ and its derivatives can be
finally expressed using the duality relation eq.~\dualinv\ in
terms of $M$ and of derivatives of the fixed-coupling dual
$\tilde\chi$ of $\gamma$: e.g. differentiating eq.~\dualinv\ with
respect to $\as$ or with respect to $M$ gives respectively
\eqnn\derduala\eqnn\derdualb
$$\eqalignno{&{\partial\over\partial\as}\gamma(\as,\tilde\chi(\as,M))
+\gamma^\prime(\as,\tilde\chi(\as,M))
{\partial\over\partial\as}{\tilde\chi}(\as,M)=0,&\derduala\cr
&\gamma^\prime(\as,\tilde\chi(\as,M)){\tilde\chi}^\prime(\as,M)=1,&\derdualb\cr}$$
and so on. Using this, the right-hand side of eq.~\nloback\ gives
us finally an expression for $\chi(\ahat,M)$ in terms of
$\tilde\chi(\ahat,M)$, i.e. for the running--coupling dual BFKL
kernel in terms of the naive dual one, which in turn can be
determined from the anomalous dimension $\gamma$ using
fixed-coupling duality eq.~\dualdef. To lowest nontrivial order
this gives the well-known~\refs{\ciafrun,\sxap} result
\eqn\derdual{\chi(\ahat,M)= \tilde\chi(\ahat,M) -
\frac{1}{2}\beta_{0}\ahat\tilde\chi(\ahat,M)
\frac{\tilde\chi''(\ahat,M)}{\tilde\chi'(\ahat,M )}+O(\ahat^2),}
where we made use of the fact that the kernels start at order
$\ahat$, so $\partial\chi(\ahat,M)/\partial\ahat
=\ahati\chi(\ahat,M)+O(\ahat)$.
Equation~\derdual\ expresses the BFKL kernel $\chi$ which is dual
to the given GLAP anomalous dimension $\gamma$, in terms of the
fixed-coupling dual $\tilde\chi$ determined from $\gamma$ using
eq.~\dualdef.

Analogously, one can construct the GLAP anomalous dimension
$\gamma$ which is dual to a given $\chi$ in terms of the
fixed-coupling dual $\tilde\gamma$ determined from $\chi$ using
eq.~\dualinv: \eqn\derdualinv{\gamma(\ahat,N)=
\tilde\gamma(\ahat,N) -
\frac{1}{2}\beta_{0}\ahat\chi(\ahat,\tilde\gamma(\ahat,N))
\frac{\chi''(\ahat,\tilde\gamma(\ahat,N))}
{\left[\chi'(\ahat,\tilde\gamma(\ahat,N))\right]^2}+O(\ahat^2),}
where $\chi(\ahat,\tilde\gamma(\ahat,N))$ can be further expressed
in terms of $\tilde\gamma(\ahat,N)$ and its derivatives
differentiating the duality relation between $\chi$ and $\tilde
\gamma$, analogously to eqs.~\derduala,\derdualb. It is important
to note that $\tilde\gamma$ and $\tilde\chi$ are not the
fixed-coupling dual of each other: rather each of them is the
fixed-coupling dual of (respectively) $\chi$ and $\gamma$, which
are related by running--coupling duality. It is sometimes
convenient~\sxsym\ (and it will be useful for our discussion of
scheme changes) to view $\tilde\chi$ as an effective $\chi$
kernel: namely, once $\gamma$ is determined from a given $\chi$
using eq.~\derdual, we also define $\tilde \chi$ which is the
fixed-coupling dual of this $\gamma$ eq.~\derdual.

In ref.~\bfcomm\ we have used this approach to derive the
running--coupling corrections to the small-$x$ resummation of the
GLAP kernel which one obtains from BFKL, i.e. essentially
eq.~\derdualinv, up to next-to-next-to-leading order. In the
sequel of this paper, we will use it to derive the running
coupling contributions to the next-to-next-to leading order BFKL
kernel obtained starting from the GLAP kernel, i.e. essentially
eq.~\derdual. Various higher-order running--coupling duality
relations are derived and collected in Appendix~A.

Because running--coupling corrections to duality are given as a
series in $\as$ of terms each of which is a function of the
fixed-coupling dual expression, also at the running--coupling level
duality maps the expansion of $\chi(\as,M)$ in powers of $\as$ at
fixed $M$ onto the expansion of $\gamma(\as,N)$ in powers of $\as$
at fixed $\as/N$, and conversely. This means that knowledge of
$\gamma$ up to next-to-next-to leading order allows us to
determine all collinear ($M\sim 0$) singular contributions to $\chi$ up to
next-to-next-to leading level. Specifically, if we expand
\eqn\chiexp{\chi(\ahat,M)=\ahat\chi_0(M)+\ahat^2\chi_1(M)+\dots,} and
then
\eqn\chimexp{\chi_i(M)=\frac{c_{i,-i-1}}{M^{i+1}}+\frac{c_{i,-i}}{M^{i}}
+\dots,} for some coefficients $c_{i,j}$, we
can determine the first three orders of the expansion
eq.~\chimexp\ of $\chi_i(M)$ for all $i$, i.e. $c_{i,j}$ for 
all $i$ and $j=-i-1,-i,-i+1$. Furthermore, the
symmetry properties of $\chi$ allow us to determine its expansion
about $M=1$ from knowledge of the coefficients of the expansion
about $M=0$. This procedure requires some care in the treatment of
the running of the coupling, which affects the way the symmetry is
realized, and it will be discussed in sect.~4. In the specific
case of $\chi_2$, knowledge of $\gamma$ up to NNLO allows us thus
to determine all the singular contributions to $\chi_2$ at $M=0$ and $M=1$,
i.e. all the collinear and ``anticollinear'' singularities respectively.

\topinsert
\vskip-0.5truecm
\vbox{
\vbox{\centerline{
\epsfxsize=8truecm
\epsfbox{chi0.eps}\hglue1truecm\epsfxsize=8truecm\epsfbox{diff0.eps}}}
\vskip0.5truecm
\vbox{\centerline{
\epsfxsize=8truecm
\epsfbox{chi1.eps}\hglue1truecm\epsfxsize=8truecm\epsfbox{diff1.eps}}}
\bigskip
\hbox{
\vbox{\footnotefont\baselineskip6pt\narrower\noindent Figure 1: 
Comparison of the exact and approximate expressions of the 
leading--order and next-to-leading--order 
contributions to the BFKL kernels $\chi_0(M)$ and $\chi_1(M)$. The
approximate expressions (given explicitly in appendix D)
are obtained by symmetrizing the third-order collinear expansion about
$M=0$ of the kernel, as discussed in section~4. The upper two plots 
are the leading-order $\chi_0$, while the lower are the 
next-to-leading-order $\chi_1$: in each case on the left we compare 
the exact (dot-dash) and approximate (solid red) 
expressions as a function of $M$, while on 
the right we show the relative difference
$(\hbox{exact}-\hbox{approximate})/\hbox{exact}$. See section~4
for a full discussion of the construction of the approximation  and
its scheme dependence.}}
\hskip1truecm}
\vskip-1.0truecm
\endinsert 
Because $\chi_0$ and $\chi_1$ are known exactly, we can test the
accuracy of the approximation to them which is obtained by
retaining only the first three terms of their expansion about
$M=0$ and $M=1$ (and subtracting the double-counting between the
two expansions). This comparison is displayed in figure~1 for the leading
and next-to-leading order contributions $\chi_0$ and $\chi_1$ to the
BFKL kernel, where it is seen that the approximation is exceedingly
good.  The percentage accuracy of the approximate expression is
better than 0.8\% at leading order and better than 1.5\% at
next-to-leading order.

This motivates us to consider the construction of a 
similar approximation to $\chi_2$. Before doing this,
however, we must turn to issues of factorization scheme
dependence.

\newsec{Factorization schemes at small $x$}
\noindent Running--coupling duality, reviewed in the previous
section, implies that, up to higher-twist corrections, the
solution to a GLAP-like equation~\tevol\ can also be obtained as
the solution to a BFKL-like equation~\xevol, and it tells us how
the evolution kernels of the two equations are related. However,
as is well known, beyond leading order, both the GLAP and BFKL
kernels are only defined up to a choice of factorization scheme.
Namely, if the normalization of $G(\xi,t)$ is redefined by a
subleading function $Z(\as,N)=1+O(\as)$, the evolution kernel
beyond leading order changes: hence, the kernel is uniquely
defined only once the normalization is fixed.

In general, of course, factorization schemes can mix the singlet
quark and gluon with each other. Here, however, we will only
consider factorization-scheme changes of the single parton
distribution $G(\xi,t)$ which enters both the GLAP and BFKL
equations: because only one eigenvector of the anomalous dimension
matrix has leading small-$x$ singularities, the BFKL equation is
one-dimensional, and mixing is irrelevant for duality. The
general matching of scheme changes which have effect at small $x$
with those which do involve operator mixing (relevant for
phenomenology) is discussed in detail in
refs.~\refs{\mom,\sxphen}. The most general factorization-scheme
change is then obtained by redefining \eqn\fact{G^\prime(t,N)=
Z_{sx}\left(\as,\frac{\as}{N}\right) Z_{lx}\left(\as,N\right)
G(t,N),} where the large-$x$ and small-$x$ scheme change functions
have respectively the form \eqnn\lxsc\eqnn\sxsc
$$\eqalignno{Z_{lx}\left(\as,N\right) &=1+\as Z^1_{lx}(N)+ O(\as^2)
&\lxsc\cr Z_{sx}\left(\as,\frac{\as}{N}\right)
&=Z^0_{sx}\left(\frac{\as}{N}\right)+\as
Z^1_{sx}\left(\frac{\as}{N}\right) +O(\as^2),  &\sxsc\cr}$$ with
the constraint that $Z^0_{sx}(0)=1$. Upon scheme change, the
leading-order contributions in the expansion of the BFKL and GLAP
kernel in powers of $\as$  remain unchanged, while higher-order
terms are modified. It is interesting to observe  that the
lowest-order nontrivial small-$x$ scheme change
$Z^0_{sx}\left(\frac{\as}{N}\right)$ (which affects the NLO BFKL
kernel $\chi_1(N)$ in eq.~\chiexp) amounts to a leading-order
redefinition of the normalization of gluon, i.e. it starts at
order $\as^0$ in the expansion of $Z$ in powers of $\as$ at fixed
$\frac{\as}{N}$~\mom.

A further related complication is due to the fact that the BFKL
equation is, in its most general form, given for a
$k_\perp$-dependent parton distribution, and  its form eq.~\xevol\
is obtained from angular averaging of this $k_\perp$-dependent
parton distribution. This angular averaging leads to an
``unintegrated'' parton distribution ${\cal G}(N,t)$, which is the
derivative of the usual parton distribution $G(N,t)$: ${\cal
G}(N,t)=\frac{d}{dt} G(N,t)$ i.e., in Mellin space,
\eqn\intuint{{\cal G}(N,M)= M G(N,M).} Because of
eq.~\ascomm, the evolution kernels $\chi(\ahat, M)$ and
$\chi_i(\ahat, M)$ for the BFKL equations satisfied by respectively
${\cal G}(M,N)$ and $ G(M,N)$ do not coincide, and are related by
\eqn\intuintker{\chi(\ahat, M) = M \chi_i(\ahat, M) M^{-1}.}
Hence, if the BFKL kernel is determined for the unintegrated gluon
distribution ${\cal G}(N,t)$, while the GLAP anomalous dimension
refers to the standard integrated quantity $G(N,t)$, there are
extra contributions to the duality relation, due to the
non-commutativity of $M$ and $\ahat$ in eq.~\intuintker.
The relation between integrated and unintegrated parton
distributions eq.~\intuint\ can be viewed as a scheme change by
using the $(N,t)$ space form of the GLAP equation~\tevol\ to
relate $\cal G$ to $G$, namely:
\eqn\intuintnt{{\cal G}(N,t)= \gamma(\as(t),N) G(N,t).}
Note that this is not a proper scheme change eq.~\sxsc\ because it
does not reduce to the identity in the limit $\as\to0$.
Note also that just as the evolution kernels for $\cal G$ and $G$ do
not coincide, similarly the anomalous dimension for their evolution
differ: in fact, from eq.~\intuint\ and the GLAP equation~\tevol\ 
one finds that $\cal G$ satisfies
\eqn\tevolunint{\frac{d}{dt}{\cal G}(N,t)=\gamma_u(\as(t),N)
{\cal G}(N,t),}
where
\eqn\gamunint{\gamma_u(\as(t),N)=\gamma(\as(t),N)+\frac{d}{dt}\ln
\left[\gamma(\as(t),N)\right].}
Henceforth, we shall denote by $\gamma$ the anomalous dimension at the
integrated level, and by $\chi$ the kernel at the unintegrated level,
and by   $\gamma_u$ and $\chi_i$ the unintegrated anomalous dimension
and integrated kernel.

Let us now come to the scheme choices which are relevant for the
explicit determination of $\gamma(\as,N)$ and $\chi(\as,M)$. The
normalization of the parton distribution which appears in the GLAP
equation~\tevol\ is fixed by the standard factorization of
collinear singularities~\cfp, and a choice of subtraction
prescription such as e.g. dimensional regularization and the \MS\
prescription. This defines anomalous dimensions in the \MS\
factorization scheme. Duality then implicitly defines a
corresponding factorization scheme for the BFKL equation. However,
the direct computation of the next-to-leading order BFKL kernel is
based on the determination~\refs{\fl,\cc,\dd} of the gluon Green
function in the high-energy limit. The extraction of the large
energy behavior of the gluon Green function, and the resummation
of its large-energy logs through the BFKL equation are based
(explicitly~\cc\ or implicitly~\refs{\fl,\dd}) on a factorization
of cross-sections in terms of a high-energy parton distribution
(the so-called $k_\perp$ factorization~\ktfact) which is
compatible with the usual collinear factorization, but differs
from it by a computable scheme change.

This is due to the fact that the usual parton distribution which
enters the collinear-factorized GLAP equation, and the gluon
density which enters the $k_\perp$-factorization formula are
normalized differently. This means that, even though the gluon
Green-function itself is computed in the \MS\ scheme, the
evolution kernel extracted from it corresponds to a scheme which
is not \MS, because it describes evolution of a quantity which
differs from the \MS\ parton distribution by a normalization
factor, i.e., it can be obtained from the \MS\ parton distribution
by a suitable scheme-change function $Z_{sx}$ eq.~\sxsc. This
scheme-change function defines the so-called $Q_0$ factorization
scheme~\ciafqz. Furthermore, the quantity which naturally enters
$k_\perp$-factorization formulae is the unintegrated parton
distribution ${\cal G}(N,t)$, so $Q_0$ scheme results are usually
given for this quantity, though, of course, $Q_0$ scheme results
can also be given for the integrated parton distribution by using
eq.~\intuintker\ and the corresponding relation eq.~\tevolunint\ 
between anomalous
dimensions.

The normalization mismatch between $k_\perp$ factorization and
collinear factorization, and thus the precise definition of the
$Q_0$ scheme, has been determined in ref.~\ktfact\ at leading
nontrivial order, i.e. at the level of $Z^0_{sx}$ eq.~\sxsc, which
affects the definition of $\chi_1$ in the expansion eq.~\chiexp\
of the BFKL kernel $\chi$, and therefor its dual GLAP anomalous
dimension $\gamma$ up to next-to-leading order (order
$\gamma_{ss}$) in the expansion of $\gamma(\as,N)$ in powers of
$\as$ at fixed $\frac{\as}{N}$:
\eqn\gamexp{\gamma(\as,N)=\gamma_s\left(\frac{\as}{N}\right)+\as
\gamma_{ss}\left(\frac{\as}{N}\right)+\dots. } The scheme change
at the next order (relevant for $\chi_2$ and $\gamma_{sss}$) has
been recently derived in ref.~\ciafscheme.\foot{Note that in
ref.~\ciafscheme\ (and elsewhere) the scheme change effected by
$Z_{sx}^0$ is referred to as a leading-log $x$ scheme change,
essentially because it is a leading-order redefinition of the
gluon normalization, whereas we will call it a NLO scheme change,
because it affects the NLO kernel $\chi_1$.}

The main result of ref.~\ciafscheme, which we shall use in what 
follows, is an expression (proven up to NNLO, but conjectured to
hold in general) which relates  the $t$ dependence of the
integrated parton distributions $G(N,t)$ (as defined in standard
collinear factorization) in the \MS\ and 
$Q_0$ scheme. This relation is expressed in terms of the BFKL kernel
for the unintegrated distribution ${\cal G}(N,t)$.
Specifically, the $t$ dependence can be written in
terms of a saddle-point evolution factor $E(t,t_0)$, a
running--coupling duality correction ${\cal N}(N,t)$, and a
normalization factor ${\cal R}(t_0)$ which is characteristic of
the way minimal subtraction with dimensional regularization is
defined by continuation of the anomalous dimensions in $d$
dimensions.

The saddle-point evolution factor is obtained by solving the
running--coupling BFKL equation for the unintegrated distribution
in the saddle-point approximation:
this can be shown~\refs{\ColKwie,\ciafrun,\sxrun} to lead to
evolution driven by the anomalous dimension 
$\tilde\gamma_u(\as(t),N)$, 
obtained from the unintegrated BFKL kernel using naive
(fixed-coupling) duality eq.~\dualdef, but with $\as=\as(t)$,
\eqn\evfactdef{E(t,t_0)=\exp\left[\int_{t_0}^t\!\tilde\gamma_u(\as(t'),N)\,dt'\right].}

The running--coupling correction to
duality discussed in the previous section can be combined with the factor
eq.~\intuintnt\ which 
relates the integrated and unintegrated
parton distributions. This gives
\eqn\rcfactrc{\eqalign{\frac{{\cal N}(N,t)}{{\cal N}(N,t_0)}&=
\frac{\gamma(\as(t_0),N)}{\gamma(\as(t),N)}
\exp\left\{\int_{t_0}^t\!\left[\gamma_u(\as(t'),N)-
\tilde\gamma_u(\as(t'),N)\right]dt'\,\right\}\cr
&=
\exp\left\{\int_{t_0}^t\!\left[\gamma(\as(t'),N)-
\tilde\gamma_u(\as(t'),N)\right]dt'\,\right\}
.\cr}}
Note that $\lim_{\as\to0}{\cal N}(N,t)=1$ so this can be viewed as a
scheme change in the proper sense.

Finally, the normalization factor $\cal R$ is due to the fact
that in the \MS\ scheme the anomalous dimension is defined as the
residue of the simple $\varepsilon$ pole in the partonic cross
section, which in turn is given by
\eqn\drev{\eqalign{\frac{\gamma(\as,N,\varepsilon)}{\beta(\as,\varepsilon)} 
&=\frac{1}{\as\varepsilon}\left(1-\frac{\beta(\as)}{\as\varepsilon}+\dots\right)\left(
\gamma(\as,N)+\varepsilon\dot{\gamma}(\as,N)+\dots
\right) \cr
&=\frac{1}{\as\varepsilon}\left(\gamma(\as,N)
-\frac{\beta(\as)}{\as}\dot{\gamma}(\as,N)+\dots
\right),}}
 where in \MS\ $\beta(\as,\varepsilon)=\as\varepsilon+\beta(\as)$ is 
the $d$-dimensional $\beta$ function, $\gamma(\as,\varepsilon)$
is the anomalous dimension obtained using duality from a
$d$-dimensional BFKL kernel, which has been expanded as
\eqn\ddad{\gamma(\as,N,\varepsilon)= \gamma(\as,N)+
\varepsilon \dot\gamma(\as,N)+O(\varepsilon^2).} It follows that the
anomalous dimension determined from duality differs from the \MS\
result through the terms beyond the first on the r.h.s. of
eq.~\drev, and thus \MS\ evolution requires an additional factor
\eqn\rfact{{{\cal R}(N,t)\over{\cal R}(N,t_0)}=
\exp\left[-\int_{t_0}^{t}\!
\beta_0\alpha(t')\dot\gamma(\alpha(t'),N)dt' + O(\as^2)\right],} 
where we have used $\beta(\as)=-\beta_0\as^2+\dots$.

Combining all these factors, the result of ref.~\ciafscheme\ takes
the form: \eqn\qzmsb{G^{Q_0}(N,t)= {\cal N} (N,t) E(t,t_0) {\cal
R}(N,t_0) G^{\ms}(N,t_0) } This equation gives the scale
dependence of the parton distribution in either scheme in terms of
a boundary condition determined in the other scheme: therefore, it
fully specifies both the relation between the two schemes, and the
scale dependence in either of them. Letting $t=t_0$ in eq.~\qzmsb\
immediately gives the relation between
$ G^{Q_0}(N,t)$ and $G^{\ms}(N,t)$, through the function
\eqn\rdef{R(N,t)\equiv {\cal N}(N,t) {\cal R}(N,t).}

The scale dependence of the parton distribution in the $Q_0$
scheme is found by keeping $t_0$ fixed in eq.~\qzmsb\ and varying
$t$: thus
\eqn\qzgevol{\eqalign{ G^{Q_0}(N,t) &= \frac{{\cal
N}(N,t)}{{\cal N}(N,t_0)} E(t,t_0)G^{Q_0}(N,t_0)\cr
 &= \exp\left[\int_{t_0}^{t}\!\gamma(\as(t'),N)\,dt'\right]
G^{Q_0}(N,t_0).\cr }}
So the integrated parton
distribution in the $Q_0$ scheme evolves with the anomalous
dimension $\gamma(N,t)$ which is related to the starting unintegrated
BFKL kernel  by running--coupling
duality combined with the transformation to the integrated level.

The scale dependence in the $\MS$ scheme is instead given by
\eqn\msbevol{\eqalign{G^{\ms}(N,t) &= \frac{{\cal R}(N,t_0)}{{\cal R}(N,t)}
E(t,t_0)G^{\ms}(N,t_0)\cr
&= \frac{{\cal R}(N,t_0)}{{\cal R}(N,t)}\exp\left[\int_{t_0}^{t}\!\tilde \gamma_u(\as(t'),N)\,dt'\right]
 G^{\ms}(N,t_0),\cr
}} namely, the integrated parton distribution in the $\MS$ scheme
evolves with an anomalous dimension which is closely related to
the fixed-coupling dual $\tilde\gamma_u(N,t)$ of the starting
(unintegrated) BFKL
kernel, and only differs from it through the scale dependence of
the ${\cal R}$ factor eq.~\rfact. In this respect the $\MS$ scheme is 
remarkably simple.

In fact it is possible, and useful, to define an
auxiliary scheme, \MSS, which differs from $\MS$ by a factor
${\cal R}(N,t)$: namely
\eqnn\mssdef 
$$\eqalignno{G^{\mss}(N,t)&= {\cal R}(N,t)
G^{\ms}(N,t)&\mssdef
}$$
In this $\MSS$ scheme the parton distribution evolves
with the naive dual anomalous dimension:
\eqn\mssevol{G^{\mss}(N,t_1)=\exp\left[\int_{t_0}^{t_1}\!\tilde
\gamma_u(\as(t),N)\,dt\right]G^{\mss}(N,t_0).}
\eqnn\mssqzgam\eqnn\mssqzchi
\topinsert
\vbox{
\vbox{\centerline{
\epsfxsize=13truecm
\epsfbox{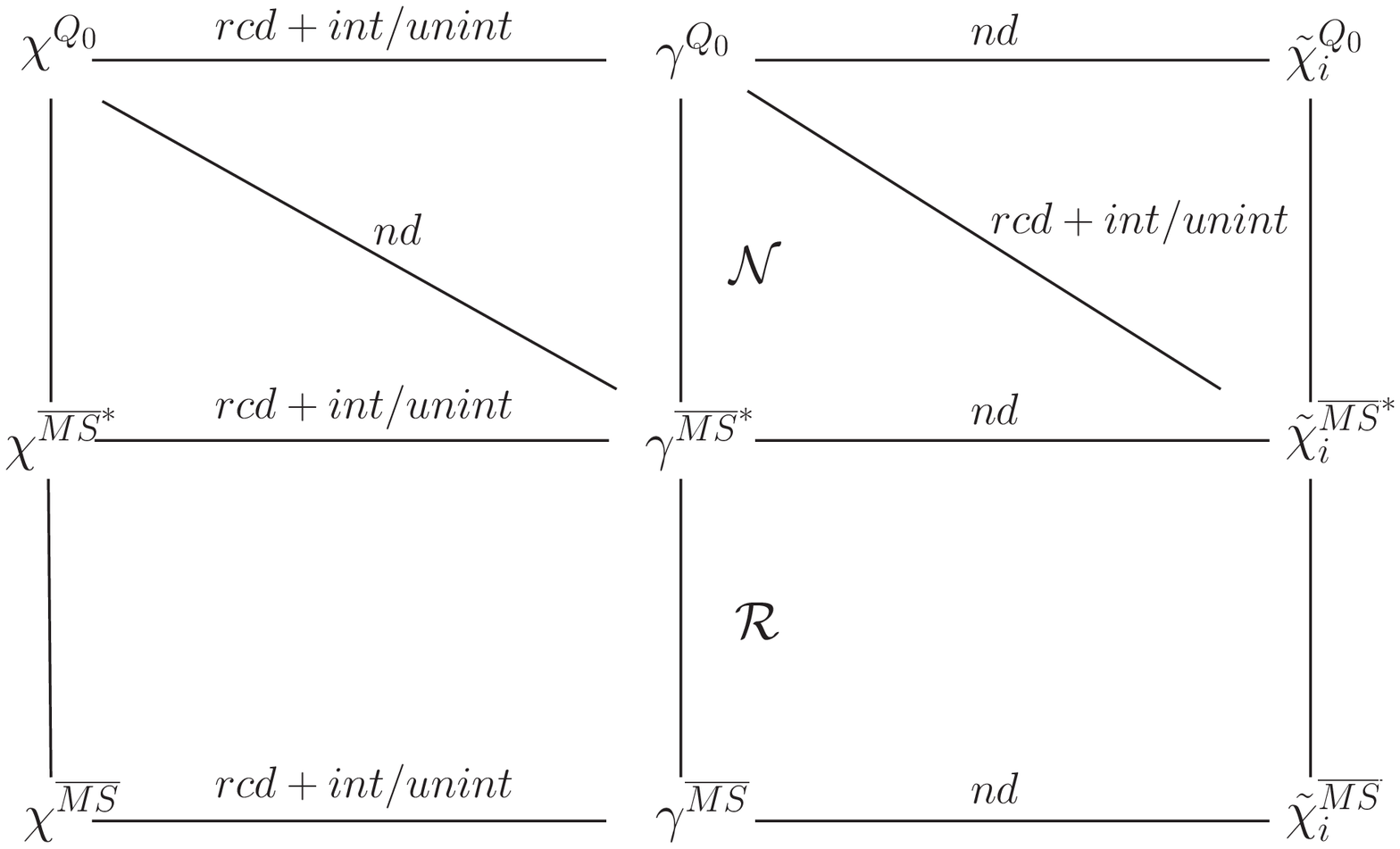}}}
\bigskip
\hbox{
\vbox{\footnotefont\baselineskip6pt\narrower\noindent Figure 2: 
Schematic relation between BFKL kernels $\chi$ and GLAP anomalous
dimensions $\gamma$ in various schemes. Horizontal lines denote
duality relations while vertical lines denote scheme
transformations. The diagonal lines express the identities
eq.~\mssqzgam\ and~\mssqzchi .}}
\hskip1truecm}
\endinsert

The relations between different quantities in various schemes,
which will be computed in the next section, are summarized in
figure~2. In the figure, horizontal lines denote duality: either
at the running--coupling level, relating $\chi$ to $\gamma$, or at
the fixed-coupling, relating $\gamma$ to $\tilde \chi$. Vertical
lines denote relations between schemes, specifically the $\Q0$,
$\MSS$ and $\MS$ schemes. Equation~\mssevol\ means that the
anomalous dimension in the $\MSS$ scheme coincides with the naive
dual of the $\Q0$  scheme BFKL kernel (at the unintegrated level):
$$\eqalignno{\tilde\gamma_u^{\q0}(\as,N)&=\gamma^{\mss}(\as,N),&\mssqzgam}$$ 
and thus, by duality, also that
$$\eqalignno{\tilde\chi_i^{\mss}(\as,M)&=\chi^{Q0}(\as,M),&\mssqzchi}$$
where  
$\tilde\chi_i^{\mss}(\as,M)$ is the naive dual of the standard GLAP
anomalous dimension in the \MSS\ scheme, while $\chi^{Q0}(\as,M)$ is
the kernel for the BFKL equation satisfied by the unintegrated parton
distribution in the \Q0\ scheme.
This further implies that if we interpret $\tilde\chi_i^{\mss}(\as,M)$ as
an operator by letting $\as\to\ahat$, and order it canonically
(i.e. in the same way as $\chi^{Q0}(\ahat,M)$), then it is related
by running--coupling duality to $\gamma^{\q0}(\as,N)$. These
relations are denoted by diagonal lines in the figure, and will
turn out to be useful in relating results obtained in various
schemes in the next section.

\newsec{The NNLO BFKL kernel}
\noindent The construction of the NNLO BFKL kernel proceeds in
three steps. First, we collect the results discussed in the
previous sections to get expressions of the BFKL kernels which
are related by either running--coupling duality ($\chi(\ahat,M)$)
or fixed-coupling duality ($\tilde\chi(\as,M)$) to a given GLAP
kernel $\gamma(\as,N)$ to NNLO in various schemes. Then, we use
the known expression of the GLAP anomalous dimensions up to
next-to-next-to leading order to determine the first three
coefficients of the expansion eq.~\chimexp\ of the leading
$\chi_0(\ahat,M)$, next-to-leading $\chi_1(\ahat,M)$, and
next-to-next-to-leading $\chi_2(\ahat,M)$ order BFKL kernels in
powers of $M$ about $M=0$. Finally, we use the underlying symmetry of
the high-energy gluon emission diagrams to determine the
corresponding coefficients of the expansion of $\chi_0(\ahat,M)$, 
$\chi_1(\ahat,M)$ and $\chi_2(\ahat,M)$, about $M=1$.

Because all duality relations are expressed in terms of naive
(fixed-coupling) dual kernels, our starting point is the
determination of the BFKL kernel $\tilde\chi(\as,M)$ which is
obtained from the NNLO GLAP anomalous dimension using the
fixed-coupling duality equation~\dualinv. Expanding
$\tilde\chi(\as,M)$ in powers of $\as$ at fixed $\as/M$ as
\eqn\chisexp{\tilde\chi(\as,M)=\tilde\chi_s(\as/M)+\as
\tilde\chi_{ss}(\as/M)+...} we have
\eqnn\chis\eqnn\chiss\eqnn\chisss
$$\eqalignno{
\gamma_0 \big( \tilde\chi_s(\as/M)\big)&= \frac{M}{\as}, &\chis\cr
\tilde\chi_{ss}(\as/M)& = -\frac{\gamma_1
\big(\chi_s(\as/M)\big)}{\gamma_0'
\big(\chi_s(\as/M)\big)}, &\chiss\cr 
\tilde\chi_{sss}(\as/M) & =
-\frac{1}{\gamma_0'\big(\chi_s(\as/M)\big)}
\Bigg[\gamma_2\big(\chi_s(\as/M)\big)
     +\gamma_1' \big(\chi_s(\as/M)\big)\chi_{ss}(\as/M) &\cr
 & \qquad
+\frac{1}{2}\gamma_0''\big(\chi_s(\as/M)\big)(\chi_{ss}(\as/M))^2\Bigg]
.&\chisss}$$

Expanding out both $\gamma_i$ and $\tilde\chi_{s^i}$, $i=1,2,3$ up to third
order in their respective arguments, eqs.~\chis-\chisss\ determine
the coefficients of these expansions in terms of each other.
Starting with the expansion of
$\gamma$\eqnn\gzexp\eqnn\goexp\eqnn\gtexp
$$\eqalignno{
\gamma_0(N)&= \frac{g_{0, -1}}{N}+g_{0, 0}+g_{0, 1} N + O(N^2),&\gzexp\cr
\gamma_1(N)&= \frac{g_{1, -1}}{N}+g_{1, 0} + O(N), &\goexp\cr
\gamma_2(N)&= \frac{g_{2, -2}}{N^2}+\frac{g_{2, -1}}{N}+ O(N^0), &\gtexp\cr
}$$
we get \eqnn\ctzexp\eqnn\ctoexp\eqnn\cttexp
$$\eqalignno{
\tilde\chi^i_0(M)&=\frac{g_{0, -1}}{M}+O(M^2), &\ctzexp\cr
\tilde\chi^i_1(M)&=\frac{g_{0, -1}g_{0, 0}}{M^2}
+\frac{g_{1,-1}}{M}+\frac{g_{2, -2}}{g_{0 -1}}+O(M),&\ctoexp\cr 
\tilde\chi^i_2(M) & =
\frac{(g_{0, -1})^2 g_{0, 1}+g_{0, -1}(g_{0, 0})^2}{M^3} 
+\frac{g_{1,0}g_{0, -1}+g_{0, 0}g_{1, -1}}{M^2} 
+\frac{g_{2,-1}}{M}+O(M^0),&\cttexp}$$
where the superscript $i$ reminds us that all these quantities are
defined at the integrated level.
 The values of the coefficients $g_{i,j}$
up to NNLO in the \MS\ scheme
can be extracted from the known GLAP anomalous
dimensions~\refs{\glap\nlo\nnlo}, and are listed in appendix B.
Using these expressions of $g_{i,j}$,
eqs.~\ctzexp-\cttexp\ give the first three terms in the expansion of
the naive dual kernel $\tilde\chi$ in the
same scheme: this corresponds to relation between $\gamma^{\ms}$ and
$\tilde\chi^{\ms}$ in figure~2.
Note that the vanishing of leading singularities of $\gamma$ at
NLO and NNLO, $g_{1, -2}=g_{2, -3}=0$, implies the well-known
vanishing of the constant and linear term in the LO BFKL
kernel~\ctzexp.

Starting from the naive dual kernel $\tilde \chi^i_k$ eq.~\ctzexp-\cttexp\
we can determine the BFKL kernel in the same approximation in various
factorization schemes.
For the sake of comparison with direct diagrammatic
computations, the  \Q0\ scheme is most relevant, because a direct
computation using minimal subtraction at the level of the BFKL
equation gives the kernel in this scheme: 
indeed, available expressions~\refs{\fl,\cc,\dd}
of the next-to-leading kernel $\chi_1$ 
provide the result in the \Q0\ scheme. Using duality, 
this kernel can be obtained from the \MS\ GLAP anomalous dimension
by exploiting eq.~\mssqzgam\ (see figure~2), which states that 
$\chi^{\q0}$ is the naive dual of the (integrated) anomalous dimension
$\gamma^{\mss}$. Hence, we can
obtain $\chi^{\q0}$ 
by first transforming the anomalous dimension to the \MSS\ scheme
through the $\cal{R}$ scheme change eq.~\rfact\
and then using naive duality eqs.~\ctzexp-\cttexp. 

The scheme change which takes the anomalous dimension to  \MSS\
can be determined, using eq.~\rfact, from the expression of
the GLAP splitting functions in $d=4-2\varepsilon$ dimensions
$\gamma(N,\varepsilon)$, as explained in Appendix~C. 
The $\MSS$ anomalous dimension is given 
by \eqnn\dgzexp\eqnn\dgoexp\eqnn\dgtexp
$$\eqalignno{
\gamma_0^{\mss}(N)&= \frac{g_{0, -1}}{N}+g_{0, 0}+g_{0, 1}N+O(N^2),&\dgzexp\cr
\gamma_1^{\mss}(N)&= \frac{g_{1, -1}}{N}+\bar{g}_{1, 0} + O(N), &\dgoexp\cr
\gamma_2^{\mss}(N)&= \frac{g_{2, -2}}{N^2}+\frac{\bar{g}_{2, -1}}{N}
+ O(N^0), &\dgtexp\cr
}$$
where all coefficients are the same as in the $\MS$ eqs.~\gzexp-\gtexp\  except
\eqn\starco{
\bar{g}_{1, 0}={g}_{1, 0}-\beta_0 \dot{g}_{0, 0} \,, \qquad
\bar{g}_{2, -1}={g}_{2, -1}-\beta_0 \dot{g}_{1, - 1}-\beta_0^2
\ddot{g}_{0, -1} . } 
The coefficients $\dot{g}_{0, 0}$ and
$\ddot{g}_{0, -1}$ are determined by the $d$-dimensional LO
splitting functions;  their values are given in Appendix~C.
The coefficient $\dot{g}_{1,- 1}$ is determined  by the
$d$--dimensional NLO GLAP kernel which is as yet unknown. This is the only 
term contributing to the singular part of the NNLO BFKL kernel that we 
have not been able to calculate explicitly: we will
estimate below the uncertainty related to our ignorance of this contribution 
to the scheme change. 

The BFKL kernel in the \Q0\ scheme is thus given in terms of the naive dual
kernel eq.~\ctzexp-\cttexp\ by
\eqnn\cqzexp\eqnn\cqoexp\eqnn\cqtexp 
$$\eqalignno{
\chi_0^{\q0}(M)&=\tilde\chi^i_0(M), &\cqzexp\cr
\chi_1^{\q0}(M)&=\tilde\chi^i_1(M)+O(M),&\cqoexp\cr 
\chi_2^{\q0}(M) & =
\frac{(g_{0, -1})^2 g_{0, 1}+g_{0, -1}(g_{0, 0})^2}{M^3} 
+\frac{\bar g_{1,0}g_{0, -1}+g_{0, 0}g_{1, -1}}{M^2} 
+\frac{\bar g_{2,-1}}{M}+O(M^0). &\cqtexp}$$

We can also determine the  BFKL kernel in the $\MS$ scheme, by
observing that it is related by  running--coupling
duality 
to the \MS\ anomalous dimension; note that this now gives the kernel
at the integrated level (figure~2 again).
Hence, we add to
eqs.~\ctzexp-\cttexp\ the running--coupling corrections discussed
in Sect.~2, and
explicitly given in appendix~A, eqs.~(A.13)-(A.15) in terms of the naive dual
expressions computed above. Defining
\eqn\deltachidef{\Delta^{\ms}\chi( M)=\chi^{i,\ms}_k(M)-\tilde\chi^i_k(M),} and
expanding about $M=0$ we
get\eqnn\dcz\eqnn\dco\eqnn\dct
$$\eqalignno{
\Delta^{\ms} \chi_0(M)&=0&\dcz\cr 
\Delta^{\ms} \chi_1(M)&=\beta_0\frac{g_{0,-1}}{M^2}&\dco\cr 
\Delta^{\ms}\chi_2(M) & =
\beta_0\left(\frac{3g_{0,0}g_{0,-1}}{M^3}+\frac{2g_{1,-1}}{M^2}
+\frac{2g_{2,-2}}{g_{0,-1}M}\right)+ 2
\beta_0^2\frac{g_{0,-1}}{M^3}+\beta_0\beta_1\frac{g_{0,-1}}{M^2}.&\dct\cr}$$

Using eqs.\dcz-\dct\ in eq.~\deltachidef\ we get 
the BFKL kernel in the \MS\ scheme
at the integrated level. The expression at the unintegrated level, to
be compared to the  unintegrated \Q0\ scheme result of eqs.~\cqzexp-\cqtexp\ 
can be obtained using using
eq.~\intuint, which implies 
\eqn\unintev {\frac{d}{d \xi} {\cal G}(\xi,M)=\left(\ahat\chi_0+
{\ahat}^2\chi_1 +{\ahat}^3\chi_2 - \left[\ahat,M\right]
\frac{\chi_0}{M}-\left[{\ahat}^2,M\right] \frac{\chi_1}{M}
 +O({\ahat}^4)\right) {\cal G}(\xi,M) ,}
so  the kernels $\chi^i_i$ at the integrated
level and $\chi_i$ at the unintegrated level are related by
\eqn\iuiexp{\eqalign{ 
\chi_0&=\chi^i_0,\cr
\chi_1&=\chi^i_1 -\beta_0\frac{\chi^i_0}{M},\cr
\chi_2&=\chi^i_2-\beta_0\beta_1\frac{\chi^i_0}{M}-2\beta_0\frac{\chi^i_1}{M}.}} 

So far, we have exploited the information from the GLAP
anomalous dimension to determine the first few terms in the
expansion of the BFKL kernel about $M=0$ (the collinear region).
However, we can further use the underlying symmetry of the BFKL
kernel to determine the corresponding terms in the expansion of
the BFKL kernel about $M=1$ (the anticollinear region). At LO and NLO 
this leads to an approximation to the BFKL kernel which, when tested on
the known LO and NLO kernels, turns out to be accurate to better
than 2\% in the whole physical region $0<M<1$, as shown in figure~1. 
Indeed, the
underlying Feynman diagrams which determine the unintegrated BFKL
kernel are symmetric upon the exchange of the incoming and
outgoing gluon. This means that the dimensionless BFKL kernel
$K(\as,k^2,Q^2)$, related to $\chi(\ahat,M)$ by Mellin
transformation \eqn\bfklkdef{\chi(\alpha_s,M)=\int_{0}^{\infty}
\frac{dQ^2}{Q^2} \left(\frac{Q^2}{k^2}\right)^{-M}
K\left(\alpha_s,k^2,Q^2\right),} is symmetric upon the interchange
of the virtualities of the incoming gluons $Q^2$ and $k^2$:
\eqn\symker{\frac{1}{Q^2}K(\as,k^2,Q^2)=\frac{1}{k^2}K(\as,Q^2,k^2).}
This  symmetry, in turn, implies that the BFKL kernel $\chi(\as,M)$ 
is symmetric upon the interchange $M\leftrightarrow 1-M$~\fl.

However, the kernel is symmetric only if one chooses a symmetric
argument for the running coupling, and if the $N$-Mellin
eq.~\nmom\ does not break the symmetry between the two scales
$Q^2$ and $\mu^2$ which enter the definition eq.~\mmom\ of 
the $M$-Mellin transform. In deep--inelastic scattering both symmetries are
broken: the argument of the running coupling is $Q^2$, and
$\xi=\ln(s/Q^2)$. Hence, the BFKL kernel which is obtained from
the GLAP anomalous dimensions incorporates these symmetry-breaking
effects. The symmetry breaking must be undone before the symmetry can be
exploited. After symmetrizing one can revert to nonsymmetric
variables and argument of the coupling in order to get an
expression of the dual $\chi$ which is accurate for all $M$.

A symmetric choice of variables is for example
$\xi=\ln(s/\sqrt{Q^2k^2})$. As is well known~\fl, the kernel
$\chi^s(\ahat, M)$ which corresponds to this choice is
related to the kernel $\chi^{\rm DIS}(\ahat, M)$  which
corresponds to DIS variables by the implicit equation
\eqn\symnonsym{ \chi^s(\ahat,M)=\chi^{\rm
DIS}\left(\ahat,M+\half\chi^s(\ahat,M)\right).}

A symmetric choice of argument for the running coupling is such
that eq.~\symker\ also holds when the argument $\mu^2$  of the
running coupling $\as(\mu^2)$ is expressed as a function of $Q^2$
and $k^2$, $\mu^2=\mu^2(Q^2,k^2)$, and 
$\mu^2(Q^2,k^2)=\mu^2(k^2,Q^2)$. Examples of symmetric choices for
the running of the coupling are $\mu^2=|Q^2-k^2|$ or
$\mu^2=\hbox{Max}(Q^2,k^2)$. Upon Mellin transformation, different
choices of arguments of the running coupling correspond to
different orderings of the running--coupling operator. For
instance, it is easy to check that the Mellin transform of
\eqn\kzero{K_0(\as,k^2,Q^2)=
\frac{C_A}{\pi}\as(Q^2)\frac{Q^2}{|Q^2-k^2|}} is $\ahat
\chi_0(M)$, where \eqn\chizerp{ \chi_0(M)=
-\frac{C_A}{\pi}\left[\psi(M)+\psi(1-M)-2 \psi(1)\right]} is  the
standard leading-order BFKL kernel, while the Mellin transform of
\eqn\kzero{K_0(\as,k^2,Q^2)=
\frac{C_A}{\pi}\as(k^2)\frac{Q^2}{|Q^2-k^2|}}
 is $ \chi_0(M)\ahat$.

Note finally that the integration which takes us from the 
unintegrated to the integrated distribution
also breaks the symmetry, as eq.~\intuintker\ explicitly shows: it
follows that 
the symmetry which holds at the unintegrated level is broken at the
integrated level.

Because the diagrammatic computation of the BFKL kernel yields 
the result in the \Q0\ scheme at the unintegrated level, once symmetric
variables and symmetric running of the coupling have been chosen
the kernel is symmetric when determined in this scheme. Whether
the symmetry is preserved or not in other schemes depends of
course on the particular scheme change. Specifically, it is easy
to see that a scheme change through ${\cal R}(N,t)$ preserves the
symmetry at leading nontrivial order (i.e. at the order of
$\chi_1$) but not beyond, essentially because it is entirely
determined by $\chi_0$ and its $d$-dimensional continuation, which
is symmetric (see appendix C), whereas a scheme change through
${\cal N}(N,t)$ breaks the symmetry already at leading nontrivial
order, because the first nontrivial running--coupling duality
correction eq.~\derdual\ is manifestly not symmetric.

As discussed already,  use of naive duality on the \MSS\ GLAP
anomalous dimension eqs.~\dgzexp-\dgtexp\ gives us the expansion
eqs.~\cqzexp-\cqtexp\ of the $Q_0$ scheme BFKL kernel for the
unintegrated parton distribution. This result is clearly in DIS
variables and can be turned into symmetric variables by expanding
out eq.~\symnonsym.  Substituting the expansion eq.~\chiexp\ of
the kernel we get \eqnn\csnsexa\eqnn\csnsexb$$\eqalignno{&
\chi^s(\ahat,M)  =
\ahat\chi_0\left(M+\half\ahat\chi_0^s(M)
+\half\ahat^2\chi_1^s(M)\right)&\cr &\qquad\qquad\qquad\qquad
+\ahat^2\chi_1\left(M+\half\ahat\chi_0^s(M)\right)+
\ahat^3\chi_2(M)+O(\ahat^4)&
 \csnsexa\cr
 &\quad\qquad\qquad=
\ahat\chi_0\left(M+\half\ahat\chi_0^s(M)\right)
+
\ahat^3\chi'_0\left(M+\half\ahat\chi_0^s(M)\right)\half\chi_1^s(M) &\cr &\qquad\qquad\qquad
 +\ahat^2\chi_1(M)
+ \ahat^3\chi'_1(M)\half\chi_0^s(M)+
\ahat^3\chi_2(M)+O(\ahat^4) .&\csnsexb\cr}$$ 
where on the right-hand side we have dropped the DIS index of eq.~\symnonsym.

The first term on the right-hand side of eq.~\csnsexb\ must be
computed by carefully keeping operator ordering into account. This
can be done by using a technique akin to that of ref.~\bfcomm,
summarized in section~2: namely, by computing
\eqn\snscalc{\eqalign{& \chi_0(M+\half\ahat\chi_0^s(M))  =
e^{\half\ahat\chi_0^s(M)\frac{d}{d\lambda}+
\half\left[M,\half\ahat\chi_0^s(M)\right]\frac{d^2}{d\lambda^2}+
\dots}\,\chi_0(M+\lambda)|_{\lambda=0} \cr &\quad =
\chi_0(M)+\half\ahat\chi_0^s(M)\chi'_0(M)+
\quarter \left[M,\ahat\chi_0^s(M)\right]\chi''_0(M)+
\eigth\ahat^2{\chi_0^s}^2(M)\chi''_0(M)+O(\ahat^3)\cr &\quad =
\chi_0(M)+\half\ahat\chi_0^s(M)\chi'_0(M)-
\quarter \beta_0\ahat^2\chi_0^s(M)\chi''_0(M)+
\eigth\ahat^2{\chi_0^s}^2(M)\chi''_0(M)+O(\ahat^3).\cr}}
Substituting this result in eq.~\csnsexb\ and collecting terms of
the same order we get\eqnn\snsfz\eqnn\snsfo\eqnn\snsft
$$\eqalignno{
\chi_0^{s}(M)&=\chi_0^{\q0}(M), &\snsfz\cr
\chi_1^{s}(M)&= \chi_1^{\q0}(M)+\half{\chi_0'}^{\q0}(M)\chi^{\q0}_0(M),
&\snsfo\cr
\chi_2^{s}(M)&=\chi^{\q0}_2(M)+\half{\chi_1'}^{\q0}(M)\chi^{\q0}_0(M)+
\half{\chi_0'}^{\q0}(M)\chi^{\q0}_1(M)\cr&+\half{{\chi_0'}^{\q0}(M)}^2\chi^{\q0}_0(M)+\eigth{\chi_0''}^{\q0}(M){\chi^{\q0}_0}^2(M)
-\quarter \beta_0{\chi_0''}^{\q0}(M)\chi^{\q0}_0(M), &\snsft\cr}$$
where on the right-hand side we have denotes the result in DIS
variables by $\chi_k^{\q0}$, since in practice we will use the
\Q0\--scheme result in DIS variables of eqs.~\cqzexp-\cqtexp.

In order to determine the constant term of $\chi_1$ and the simple 
pole of $\chi_2$ in symmetric variables we have substituted the expansion 
of $\chi_0$ up to $O(M^2)$. In principle the linear term of $\chi_1$ is 
needed too, but its dependence cancels out in the expression for $\chi_2$.
Using the expressions eq.~\cqzexp-\cqtexp\ for the unintegrated
\Q0--scheme kernels $\chi_i^{\q0}$ 
we finally get \eqnn\cfz\eqnn\cfo\eqnn\cft
$$\eqalignno{\chi_0^{s}(M) & =  \frac{g_{0,-1}}{M} + O(M^2) \,, &\cfz\cr
\chi_1^{s}(M) & =  -\frac{(g_{0,-1})^2}{2 M^3}+
                    \frac{g_{0,0} g_{0,-1}}{M^2}+
             \frac{g_{1,-1}}{M}+
                \frac{g_{2,-2}}{g_{0,-1}}+ g_{0,-1}^2 \zeta(3)  +O(M) \,,   &\cfo\cr
\chi_2^{s}(M) & =  \frac{(g_{0,-1})^3}{2 M^5}-
                   \frac{3 g_{0,0} (g_{0,-1})^2+\beta_0 (g_{0,-1})^2}{2 M^4}+
             \frac{(g_{0,0})^2 g_{0,-1}
+g_{0,1} (g_{0,-1})^2-g_{0,-1} g_{1,-1}}{M^3} \cr
  & +\frac{-\half g_{2,-2}+g_{0,0} g_{1,-1}+g_{0,-1} \bar{g}_{1,0}}{M^2}  
  +\frac{\bar{g}_{2,-1}-2 \beta_0 (g_{0,-1})^2 \zeta(3)}{M} +O(M^0)  
\,. &\cft}$$

We can now exploit the symmetry of the kernel with symmetric
coupling, which implies that $\chi(\ahat,M)$ must admit an
expansion of the form\eqnn\chisyexpa\eqnn\chisyexpb
$$\eqalignno{\chi(\ahat,M)&=\ahat\chi^{s}_0(M)+\ahat^2\chi^{s}_1(M)+\ahat^3\chi^{s}_2(M)+O(\ahat^3)&\chisyexpa\cr
&=\chi^{sym}_0(\ahat,M)+\chi^{sym}_1(\ahat,M)+\chi^{sym}_2(\ahat,M)+O(\ahat^3)&\chisyexpb}$$
where $\chi_i^{sym}(\ahat,M)$ are the symmetrized functions
\eqnn\csz\eqnn\cso\eqnn\cst
$$\eqalignno{\chi_0^{sym}(\ahat,M)&=c_{0,-1} \left[\ahat
\frac{1}{M}+\frac{1}{1-M}\ahat\right]+ \ahat c_{0,0} + c_{0,1} \left[\ahat
M +(1-M)\ahat\right]&\cr&\qquad+c_{0,2}\left[\ahat  M^2
+(1-M)^2\ahat\right]+O(M^3),&\csz\cr 
\chi_1^{sym}(\ahat,M)&=c_{1,-3}
\left[\ahat^2
\frac{1}{M^3}+\frac{1}{(1-M)^3}\ahat^2\right]+c_{1,-2}
\left[\ahat^2 \frac{1}{M^2}+\frac{1}{(1-M)^2}\ahat^2\right] &\cr
&\qquad+ c_{1,-1} \left[\ahat^2
\frac{1}{M}+\frac{1}{1-M}\ahat^2\right] + \ahat c_{1,0} + c_{1,1}
\left[\ahat^2 M +(1-M)\ahat^2\right]&\cr &\qquad+O(M^2),&\cso\cr
\chi_2^{sym}(\ahat,M)&=\sum_{j=1,5} c_{2,-j} \left[\ahat^3
\frac{1}{M^j}+\frac{1}{(1-M)^j}\ahat^3\right]
+O(M^0). &\cst\cr
}$$

Given the expression of $\chi_i^{sym}$ eq.~\chisyexpb\ it is
straightforward to determine the symmetrized kernel when $\ahat$ is
``canonically''  ordered to the left,  which corresponds to the choice
of argument of the running coupling $\as=\as(Q^2)$:
\eqnn\cantosz\eqnn\cantoso\eqnn\cantost
$$\eqalignno{\ahat\bar\chi^{sym}_0(M)&=\chi_0^{sym}(\ahat,M),&\cantosz\cr
\ahat^2\bar\chi^{sym}_1(M)&=\chi_1^{sym}(\ahat,M) - \ahat^2 \beta_0
\frac{c_{0,-1}}{(1-M)^2} 
+ \beta_0\ahat^2 (c_{0,1}+2 c_{0,2})+O(M), & \cantoso\cr
\ahat^3\bar\chi^{sym}_2(M)&=\chi_2^{sym}(\ahat,M) 
- \ahat^3 \beta_0 \beta_1 \frac{c_{0,-1}}{(1-M)^2}
+ 2 \ahat^3 \beta_0^2 \frac{c_{0,-1}}{(1-M)^3} \cr
 & -2 \ahat^3 \beta_0 \frac{c_{1,-1}}{(1-M)^2}
- 4 \ahat^3 \beta_0 \frac{c_{1-2}}{(1-M)^3}
-8\ahat^3 \beta_0 \frac{c_{1-3}}{(1-M)^4}+O(M^0).
&\cantost\cr}$$
In $\bar\chi^{sym}_i$ the symmetry is broken by the running of the
coupling only.

\topinsert
\vbox{
\centerline{
\epsfxsize=12truecm
\epsfbox{chi2.eps}}
\bigskip
\hbox{
\vbox{\footnotefont\baselineskip6pt\narrower\noindent Figure 3: The
approximate expression of the NNLO contribution to the BFKL kernel for
the unintegrated distribution
eq.~\cantost\ in the \Q0\ scheme, symmetric variables, $\as(Q^2)$. See
appendix~D for the values of all coefficients. The uncertainty band is
obtained varying the unknown scheme---fixing coefficient $-5\leq\dot
g_{1,-1}\leq5$ (top to bottom).
}}}
\vskip1.5truecm
\vbox{
\centerline{
\epsfxsize=12truecm
\epsfbox{nnlodiff.eps}}
\bigskip
\hbox{
\vbox{\footnotefont\baselineskip6pt\narrower\noindent Figure 4: The
relative uncertainty in the NNLO contribution to the BFKL kernel for
the unintegrated distribution shown in figure~3, due to the uncertainty 
in the unknown scheme---fixing coefficient $\dot g_{1,-1}$, here taking the values
$-5,-1,1,5$ (top to bottom).
}}}
\endinsert 

We can finally determine all coefficients $c_{ij}$ in eqs.~\csz-\cst\
by expanding the symmetrized kernel eqs.~\cantosz-\cantost\
in Laurent series about $M=0$ and equating to the
expansion of the unsymmetrized kernels $\chi_i^s$ eqs.~\cfz-\cft,
which is accurate to the stated power of $M$.
Because the anticollinear terms with poles at $M=1$ in eqs.~\csz-\cst\ are regular in
$M=0$, the symmetrized $\chi_i^{sym}$ have the same
$M=0$ poles as their unsymmetrized counterparts
$\chi_i^s$, and their coefficients can be read off
eq.~\cfz-\cft.
However, the anticollinear terms do contribute to all
regular contributions in the expansion of  $\chi_i^s$ about
$M=0$. This is why higher--order regular terms must be included in the
right-hand side of eqs.~\csz,\cso: specifically, symmetric
terms up to
$O(M^2)$ must be included in $\chi^s_0(M)$ in order for its expansion to coincide with
that of $\chi^{sym}_0(M)$ up to and including $O(M)$, and terms 
up to and including 
$O(M)$ in $\chi^s_1(M)$ in order for its expansion to coincide with
that of $\chi^{sym}_1(M)$ up to $O(M^0)$.
No addition is necessary for  $\chi_2$ because the known
coefficients in its expansion  about $M=0$ are all singular.

Summarizing, all singular coefficients in eqs.~\csz-\cst\ can be read off
eq.~\cfz-\cft, while for the nonsingular ones we get
\eqn\ccoeffs{\eqalign{
c_{0, 0}  & = - \smallfrac{3}{2}g_{0,-1},\qquad
c_{0, 1}   = 0,\qquad
c_{0, 2}   = \half g_{0,-1}, \cr
c_{1,  0} & = \smallfrac{g_{2,-2}}{g_{0,-1}}+ (g_{0,-1})^2 \zeta(3),\qquad
c_{1,  1}  = \half (g_{0,-1})^2 - g_{0,-1} g_{0,0}- g_{1,-1}.}}
Using these results in eqs.~\cantosz-\cantost\ we get our
approximate expression for the BFKL kernel up to NNLO, 
at the unintegrated level in the \Q0\ scheme in
symmetric variables, with $\as=\as(Q^2)$. The LO kernel of course does not
depend on either scheme, the choice of variables, or the running of
the coupling.
The NLO kernel  corresponds 
to the widely used form of
the kernel as given in ref.~\fl, eq.~(14) of that reference. Indeed, 
it can be straightforwardly checked that the Laurent
expansion of eq.~\cantoso\ coincides with the result of ref.~\fl\ up
to and including $O(M^0)$. The NNLO kernel eq.~\cantost\ is a new result.
The expressions for the kernel in DIS variables can be obtained
straightforwardly from eqs.~\cantosz-\cantost\ by inverting
eqs.~\snsfo-\snsft, and the kernel at the integrated level is found
using eq.~\iuiexp. The  \MS\ scheme expressions are found using
eqs.~\dco-\dct\ in eq.~\deltachidef. 

\topinsert
\vbox{
\centerline{
\epsfxsize=12truecm
\epsfbox{nnlo.eps}}
\vskip0.5truecm
\bigskip
\hbox{
\vbox{\footnotefont\baselineskip6pt\narrower\noindent Figure 5: The
full BFKL kernel at leading, next-to-leading, and next-to-next-to
leading order obtained combining tyhe known expressions  for the
LO and NLO contributions (as shown in figure 2) and our approximate
expression (with $\dot g_{1,-1}=0$) for NNLO  (as shown in figure~3), 
with $\ahat\to0.2$. All expressions are in the \Q0\ factorization
scheme, at the unintegrated level with
symmetric variables, and $\as=\as(Q^2)$. The symmetry  about $M=\half$
is only broken
by the argument of the running coupling.
}}}
\endinsert 

The full expression for the kernel up to NNLO  in the \Q0\ scheme
is given in Appendix~D, both in symmetric and DIS variables, at the
unintegrated and integrated level.  The approximate expressions
of the LO and NLO kernels   compared to the
exact kernels in figure~1 correspond to the expressions
eq.~\cantosz-\cantoso\ (\Q0\ scheme, unintegrated, symmetric
variables, $\as(Q^2)$). The approximate expression of $\chi_2$ is
displayed in figure~3, in the \Q0\ scheme, in symmetric variables,
at the unintegrated level and with the canonical argument of the coupling 
$\as(Q^2)$. As  discussed above, eq.~\starco\ (see also
Appendix C) one of the coefficients which determine the
next-to-next-to-leading order scheme change between the \MS\ and \MSS\
(and thus \Q0), namely $\dot g_{1,-1}$, is unknown. This coefficient affects the simple
(sub-subleading) poles, and has therefore a moderate impact. Noting
that all the scheme-change coefficients (see Appendix~C,
eq.~(C.7)) are of order one or smaller, and indeed all coefficients
$g_{i,j}$ (see Appendix~B,
eq.~(B.2)) are at most of order of a few, we estimate the uncertainty
related to this coefficient by varying $-5\le \dot{g}_{1,-1}\le5$. The
corresponding uncertainty is displayed in
figure~3, and the relative uncertainty in figure~4: it is seen 
to be similar to the uncertainty of a few
percent that we expect (on the basis of the LO and NLO results of
figure~1) to affect our approximate form of $\chi_2$.

\topinsert
\vbox{
\centerline{
\epsfxsize=12truecm
\epsfbox{intercept.eps}}
\bigskip
\hbox{
\vbox{\footnotefont\baselineskip6pt\narrower\noindent Figure 6: The
BFKL intercept $c(\as)$ at leading, next-to-leading, and 
next-to-next-to-leading order, obtained from the $\chi$ as shown in figure 5, 
with a symmetric argument for the running coupling, and 
evaluated at $M=\half$.
}}}
\endinsert 

In figure~5 we display the full NNLO BFKL kernel
$\chi(\ahat,M)=\ahat \chi_0(M)+\ahat^2\chi_1(M)+\ahat^3\chi_2(M)$,
with the same scheme and variable choices
using the exact expressions up to NLO and the approximate expression
to NNLO, with $\ahat\to 0.2$. The slow convergence properties of the
expansion of the BFKL kernel, driven by the increasingly dominant 
collinear and anti-collinear singularities at $M=0$ and $M=1$, are 
very apparent in this figure. Although the $\chi_2$ contribution 
restores the minimum near $M=\half$, the convergence of the expansion 
in the vicinity of the minimum is still rather slow.

This is made more explicit in figure~6, where we plot the intercept
$c(\as)$ as a function of $\as$. This is calculated by using the same
expression for $\chi$ as in figure~5, but with the coupling chosen to 
be symmetric, so that all the curves are symmetric about $M=\half$, and 
then defining $c(\as) = \chi(\as,\half)$, ie the value of $\chi$ the 
stationary point. While the fixed order perturbation series is 
clearly good for very small $\as$, say $\as \lsim 0.05$, with $\as\sim 0.1$
there are signs that the series has yet to converge. For yet larger
values of $\as$ (as would be appropriate for phenomenological studies) the
results from the fixed order series are clearly not very useful, and a 
resummation of collinear and anti-collinear singularities along 
the lines discussed in Ref.~\refs{\salam,\ciafres,\sxres,\sxsym} 
becomes necessary.

\newsec{Outlook}

In this paper we have presented an approximate determination of the
NNLO contribution to the BFKL kernel. In the process, we have
provided a full treatment to this order of various  issues which affect the
determination of the BFKL kernel: the relation between the \MS\ and \Q0\
factorization schemes, the duality relations which connect the BFKL
kernel
to the GLAP anomalous dimension, specifically in the presence of running
coupling, the choice of kinematic variables in the definition of the
BFKL kernel, the relation between the form of the BFKL kernel and the
argument of the running coupling, and the relation between BFKL kernels for
integrated and unintegrated parton distributions. All these issues
become rather nontrivial to next-to-next-to leading order, and require
full control of factorization scheme and running coupling.

Because the perturbative
expansion of the BFKL kernel in both the collinear and anticollinear
regions is alternating in sign, a knowledge of NNLO corrections is
necessary for an accurate assessment of the uncertainty involved in a
fixed--order determination of the kernel: indeed, whereas the qualitative
features of the NLO kernel are completely different from those of the
LO, the NNLO result is qualitatively similar, though we have shown
that it is quantitatively not so reliable because of the slow
convergence of the perturbative expansion, even in the central region 
away from the singularities.

Fixed--order BFKL kernels have been widely used recently in studies of
nonlinear (saturation) corrections to the BFKL equation and their
phenomenological implications for RHIC and the LHC (see
e.g.~\enterrev\ and ref. therein). Also, they are the foundation of
numerical approaches to the BFKL equation (see~\andersen\ and
ref. therein),
which in turn are relevant for Monte Carlo simulations 
(see e.g. ref.~\lund\ and ref. therein). 
Because of the slow convergence of the perturbative expansion, the
determination of the NNLO BFKL kernel presented
here is useful in assessing the reliability of these calculations. 
Finally, the
approximate form of the LO and NLO kernels given here are extremely
accurate while having a very simple analytic form, and are thus
amenable to simple numerical and phenomenological implementations. 

\noindent{\bf Acknowledgement:} We thank T.~Binoth, V.~del~Duca,
G.~Ridolfi and A.~Vicini for various discussions during the course of
this work.  S.~M. is
supported through a SUPA graduate studentship. 
S.~F. was partly supported by a PPARC visiting fellowship
and a PRIN2004 grant (Italy). 
\vfill\eject

\noindent
\appendix{A}{Higher--order duality}
Higher-order duality relations can be obtained by pursuing to
higher orders the expansion of the Baker-Campbell-Hausdorff
equation for a pair of non-commuting operators $\hat p$, $\hat q$
which act in the same way on a state $G$, according to eq.~\opeq.
To fifth order we get \eqn\bchfive{\eqalign{ f(\hat{q})G(N,M)&
=\bigg[
f(\hat{p})-\half f(\hat{p})^{''}\left[\hat{p},\hat{q}\right]
+\smallfrac{1}{6}f(\hat{p})^{'''}\left[\hat{q},\left[\hat{q},\hat{p}\right]\right]+
\cr &\qquad
+\third f(\hat{p})^{'''}\left[\hat{p},\left[\hat{p},\hat{q}\right]\right]
+\smallfrac{1}{24}f(\hat{p})^{IV}\left[\hat{q},\left[\hat{q},\left[\hat{q},\hat{p}\right]\right]\right]+
\cr &\qquad
+\eigth f(\hat{p})^{IV}\left[\hat{q},\left[\hat{p},\left[\hat{p},\hat{q}\right]\right]\right]
-\eigth f(\hat{p})^{IV}\left[\hat{p},\left[\hat{p},\left[\hat{p},\hat{q}\right]\right]\right]+
\cr &\qquad
+\eigth f(\hat{p})^{IV}{\left[\hat{p},\hat{q}\right]}^2
-\smallfrac{1}{24}f(\hat{p})^{V}\left[\hat{p},\hat{q}\right]\left[\hat{q},\left[\hat{q},\hat{p}\right]\right]+
\cr &\qquad
-\smallfrac{1}{24}f(\hat{p})^{V}\left[\hat{q},\left[\hat{q},\hat{p}\right]\right]\left[\hat{p},\hat{q}\right]
-\smallfrac{1}{12}f(\hat{p})^{V}\left[\hat{p},\hat{q}\right]\left[\hat{p},\left[\hat{p},\hat{q}\right]\right]+
\cr &\qquad
-\smallfrac{1}{12}f(\hat{p})^{V}\left[\hat{p},\left[\hat{p},\hat{q}\right]\right]\left[\hat{p},\hat{q}\right]
-\smallfrac{1}{48}f(\hat{p})^{VI}{\left[\hat{p},\hat{q}\right]}^3+O({\ahat}^4)\bigg]G(N,M)
\ .}}

We can now obtain higher-order generalizations of the BFKL-like
equation obtained starting from a GLAP equation by a suitable
identification of $\hat p$ and $\hat q$. Because, as well-known
(see e.g.~\bfcomm) at the fixed-coupling level the dual of the
expansion of $\chi(\as,M)$ in powers of $\as$ at fixed $M$ is the
expansion of $\gamma(\as,N)$ in powers of $\as$ at fixed $\as/N$,
it is convenient to expand \eqn\gamsexp{\gamma(\hat{\alpha}_s,N
{\hat{\alpha}_s}^{-1})=\gamma_s(N
{\hat{\alpha}_s}^{-1})+\hat{\alpha}_s\gamma_{ss}(N
{\hat{\alpha}_s}^{-1})+\dots} We then perform the identification
eq.~\opidinv\ of $\hat p$ and $\hat q$ and use eq.~\bchfive\ with
$f(\hat p)=\bar \chi(\ahat,\hat p)$, where
\eqn\bcdef{\bar\chi(\hat{\alpha},\gamma(\hat{\alpha},N
{\hat{\alpha}}^{-1}))=N {\hat{\alpha}}^{-1}.} Namely, $\bar
\chi(\ahat,M)=\ahati\tilde\chi(\ahat,M) $, where $\tilde \chi$ is
the naive (fixed-coupling) dual of $\gamma$, so that \eqn\chiexp
{\bar\chi(\hat{\alpha},M)=\tilde\chi_0(M)+\hat{\alpha}\tilde\chi_1(M)+{\hat{\alpha}}^2\tilde\chi_2(M)+...\,.}

Using eq.~\bchfive\ we then get
\eqn\rcdualin{\eqalign{
N {\hat{\alpha}}^{-1}
G(N,M)&=\Big\{\bar{\chi}({\hat{\alpha},M})-\half
[M,\hat{\gamma}]\bar{\chi}^{''}(\hat{\alpha},M)
-\frac{1}{6}
[M,[M,\hat{\gamma}]] \bar{\chi}^{'''}(\hat{\alpha},M)+
\cr
&\qquad+\frac{1}{6}
[\hat{\gamma},[\hat{\gamma},M]]
\bar{\chi}^{'''}(\hat{\alpha},M)+
\eigth{[M,\hat{\gamma}]}^2
\bar{\chi}^{IV}(\hat{\alpha},M)\Big\} G(N,M)\, ,}}
where primes denote derivatives with respect to $M$, and
we define $\hat\gamma\equiv \gamma(\hat{\alpha}_s,N
{\hat{\alpha}_s}^{-1})$. 

All commutators can now be determined explicitly using the expression
\eqn\ahatnnlo{
{\hat{\alpha}_s}^{-1}=\frac{1}{\alpha_s}-\beta_0\frac{\partial}{\partial M}+\beta_1\left(-\alpha_s\beta_0\frac{\partial}{\partial M}-
\half(\alpha_s\beta_0)^2\frac{\partial^2}{\partial M^2}\right)
+O(\alpha_s^3)}
of the running coupling at the operator level:
\eqnn\coma\eqnn\comb\eqnn\comc\eqnn\comd
$$\eqalignno{
&[\hat{\gamma},M]  =
-(N\beta_{0}+N\beta_{0}\hat{\alpha}_s \beta_{1})\frac{\partial\gamma(\hat{\alpha}_s,
N\hat{\alpha}_s^{-1})}{\partial N\hat{\alpha}_s^{-1}}+
&\cr
&\qquad+\beta_{0}\hat{\alpha}_s^{2}\frac{\partial\gamma(\hat{\alpha}_s,
N\hat{\alpha}_s^{-1})}{\partial \hat{\alpha}_s}+
O(\hat{\alpha}_s^{3}), &\coma\cr
&[\hat{\gamma},[\hat{\gamma},M]] = O(\hat{\alpha}_s^{3})&\comb\cr
&[M,[M,\hat{\gamma}]] =
(N\beta_{0})^{2}\frac{\partial^{2}\gamma(\hat{\alpha}_s,
N\hat{\alpha}_s^{-1})}{\partial(N\hat{\alpha}_s^{-1})^{2}} +
O(\hat{\alpha}_s^{3}),&\comc\cr
&\left([\hat{\gamma},M]\right)^{2} = (N\beta_{0})^{2} \left(
\frac{\partial \gamma(\hat{\alpha}_s, N\hat{\alpha}_s^{-1})}{\partial
N\hat{\alpha}_s^{-1}}\right)^{2} + O(\hat{\alpha}_s^{3}).&\comd}$$

Substituting the commutators \coma-\comd\ in eq.~\rcdualin\ and
back-substituting order by order the low-order expansion of the
enduing equation in the higher-order terms one may remove all the
$N$ dependence from the right-hand side of eq.~\rcdualin, with the
result: \eqn\rcdualout{\eqalign{ N{\ahat}^{-1} G & =
\Big\{\bar{\chi}(\ahat,M) - \half\ahat\beta_{0}
\frac{\bar{\chi}(\ahat,M)\bar{\chi}''(\ahat,M)}{\bar{\chi}^{'}(\ahat,\hat{\gamma})}
+ \cr &\qquad+{\ahat}^2
\Big[\quarter {\beta_{0}}^2\bar{\chi}(\ahat,\hat{\gamma}){\frac{\bar{\chi}^{''}(\ahat,\hat{\gamma})}
{\bar{\chi}^{'}(\ahat,\hat{\gamma})}}^2-\half\beta_{0}\beta_{1}
\frac{\bar{\chi}(\ahat,M)\bar{\chi}''(\ahat,M)}{\bar{\chi}^{'}(\ahat,\hat{\gamma})}+
\cr &\qquad
+{\beta_{0}}^2\frac{{\bar{\chi}(\ahat,M)}^2}{24{\bar{\chi}^{'}(\ahat,M)}^4}\Big(12
(\bar{\chi}^{''}(\ahat,M))^3+ \cr &\qquad -7
\bar{\chi}^{'}(\ahat,M) \bar{\chi}^{''}(\ahat,M)
\bar{\chi}^{'''}(\ahat,M)+3 (\bar{\chi}^{'}(\ahat,M))^2
\bar{\chi}^{IV}(\ahat,M)\Big)+ \cr &\qquad-\half
\beta_{0}\frac{\partial\bar{\chi}(\ahat, M)}{\partial
\ahat}\bar{\chi}''(\ahat,M)\Big]\Big\}G\,. }}

Identifying the term in curly brackets in eq.~\rcdualout\ with the
BFKL kernel \eqn\chibexp
{\chi(\ahat,M)=\ahat\chi_0(M)+\ahat\chi_1(M)+{\ahat}^2\chi_2(M)+...\,.}
and expanding $\bar\chi$ as in eq.~\chiexp, eq~\rcdualout\ gives
an order-by-order expression of the running--coupling dual in terms
of the naive dual. Up to NNLO we get \eqnn\cgz\eqnn\cgo\eqnn\cgt
$$\eqalignno{
{\chi}_0 & =  \tilde\chi_0 &\cgz\cr
{\chi}_1 & =  {\tilde\chi}_1 - \half\beta_0
\frac{{\tilde\chi}_0 {\tilde\chi}_0''}{{\tilde\chi}_0'} &\cgo\cr
\chi_2 & =  \tilde{\chi}_2
-\half\beta_0 \beta_1 \frac{\tilde{\chi}_0 \tilde{\chi}_0''}{\tilde{\chi}_0'}+
\cr
&\quad+\frac{1}{24}\beta_0^2
\frac{(\tilde{\chi}_0)^2}{(\tilde{\chi}_0')^4}\left(12\left(\tilde{\chi}_0''\right)^3 -
14\tilde{\chi}_0'\tilde{\chi}_0''\tilde{\chi}_0'''+3\left(\tilde{\chi}_0'\right)^2
{\tilde{\chi^{IV}}_0}\right)+
&\cr
&\quad-\half\beta_0 \frac{\tilde{\chi}_0 \tilde{\chi}_1''}{\tilde{\chi}_0'}- \beta_0
\frac{\tilde{\chi}_1 \tilde{\chi}_0''}{\tilde{\chi}_0'}+ \half\beta_0 \frac{\tilde{\chi}_0
\tilde{\chi}_0''\tilde{\chi}_1'}{\left(\tilde{\chi}_0'\right)^2}&\cr
&\quad+\quarter  \beta_0^2 \frac{\tilde{\chi}_0}{(\tilde{\chi}_0^{'})^2}\left(2
\tilde{\chi}_0^{'} \tilde{\chi}_0^{'''}- (\tilde{\chi}_0^{''})^2
\right), &\cgt}$$
where all $\chi_i$ and $\tilde \chi_i$ are functions of $M$ and the
prime denotes differentiation with respect to $M$.

By inverting this relation (i.e. expressing order by order $\tilde
\chi$ in terms of $\chi$), this result can be used to determine
the running--coupling corrections to the anomalous
dimension $\gamma$ determined  from a given BFKL kernel $\chi$. These
corrections up to NNLO were first derived in ref.~\mf, and then
reproduced more recently in refs.~{\ciafscheme,\bfcomm}. However
in both cases only the leading-order kernel $\chi_0$ and
leading-order running of the coupling were included. For
completeness, we derive them here consistently including the
running of the coupling up to NLO.

Namely, we start from a given BFKL kernel $\chi(\ahat,N)$ and
determine first the GLAP anomalous dimension $\tilde
\gamma(\as,N\as^{-1})$ which is obtained from it  using naive
duality, which we then expand according to eq.~\gamsexp. We wish
to determine the anomalous dimension $\gamma(\ahat,N\ahat^{-1})$
which is related through running--coupling duality to the starting
$\chi$. We observe that the naive-duality relations between $\chi$
and $\tilde\gamma$ and between $\tilde\chi$ and $\gamma$ imply
that \eqn\dualdirinv{N/\as=\chi(\as,\tilde\gamma(\as,N/\as))=
\tilde\chi(\as,\gamma(\as,N/\as)).} Note that in eq.~\dualdirinv\
it is immaterial whether $\as$ is considered to be an operator or
not, because $[\ahat,N]=0$ anyway. Defining
\eqn\cdcgdg{\eqalign{\Delta
\chi(\ahat,M)&=\tilde{\chi}(\ahat,M)-\chi(\ahat,M)\cr \Delta
\gamma(\ahat,N\ahat^{-1})&=\tilde\gamma(\ahat,N\ahat^{-1})
-\gamma(\ahat,N\ahat^{-1}),\cr}} and expanding eq.~\dualdirinv\ in
powers of $\as$ we get \eqn\dcfg{\eqalign{ \Delta \chi_1&=-\chi_0'
\Delta \gamma_{ss}\cr \Delta \chi_2&=-\chi_0' \Delta
\gamma_{sss}-\chi_1 {\Delta \gamma_{ss}}'-\chi_1' \Delta
\gamma_{ss}+\half\chi_0'' {\Delta \gamma_{ss}}^2+\chi_0'
{\Delta \gamma_{ss}}'\Delta \gamma_{ss},\cr}} where
$\Delta\chi_i(M)$ and $\Delta\gamma_i(N\ahat^{-1})$ are
respectively the coefficients of the expansion of
$\Delta\chi(\ahat,M)$ and $\Delta \gamma(\ahat,N\ahat^{-1})$ in
powers of $\ahat$, and on the left-hand side all $\Delta\chi_i$
are evaluated as functions
$\Delta\chi_i(\tilde\gamma(\as,N\as^{-1}))$.

Furthermore, inverting eq.~\cgt\ we can determine the coefficients
$\Delta\chi_i$ of the expansion of $\Delta\chi(\as,M)$ in powers
of $\as$ in terms of $\chi$ (all $\chi_i$ functions of $M$):
\eqn\dcfc {\eqalign{\Delta \chi_1&=\half\beta_0 \frac{\chi_0
\chi_0''}{\chi_0'} \cr \Delta \chi_2&=\half\beta_0 \beta_1
\frac{\chi_0 \chi_0''}{\chi_0'}+ \half\beta_0 \frac{\chi_0
\chi_1''}{\chi_0'}+ \beta_0 \frac{\chi_1 \chi_0''}{\chi_0'}-
\half\beta_0 \frac{\chi_0
\chi_0''\chi_1'}{\left(\chi_0'\right)^2}\cr&
\quad+\frac{1}{24}\beta_0^2 \frac{(\chi_0)^2}{(\chi_0')^4}\left(6
\left(\chi_0''\right)^3 -
10\chi_0'\chi_0''\chi_0'''+3\left(\chi_0'\right)^2
{\chi_0}^{IV}\right) +\quarter  \beta_0^2
\frac{\chi_0}{(\chi_0^{'})^2}{\chi_0^{''}}^2.}}

Equating the right-hand side of each of eqs.~\dcfg\ to the
corresponding equations~\dcfc\ we determine finally \eqn\dgss{
\Delta \gamma_{ss}=-\half\beta_0  \frac{\chi_0
\chi_0''}{\chi_0'^2}\Big|_{{M = \tilde{\gamma}_s}}} and \eqn\dgsss
{\Delta \gamma_{sss}=\Delta \gamma_{sss}^{(0)}+\Delta
\gamma_{sss}^{(1)}+ \beta_1 \Delta \gamma_{ss},} where we have
defined \eqn\dgsssz{\Delta
\gamma_{sss}^{(0)}=-\frac{1}{24}\beta_0^2
\frac{(\chi_0)^2}{(\chi_0')^5}\left(15\left(\chi_0''\right)^3 -
16\chi_0'\chi_0''\chi_0'''+3\left(\chi_0'\right)^2
{\chi_0}^{IV}\right)\Big|_{M = \tilde{\gamma}_s}}
and  
\eqn\dgssszz{\Delta\gamma_{sss}^{(1)}=\left(\half \beta_0 \chi_0 \chi_0'
{\tilde{\gamma}_{ss}}'' +\half \beta_0 \chi_0 \chi_1'
\tilde{\gamma}_{s}''+\half \beta_0 \chi_0' \chi_1
\tilde{\gamma}_{s}''+ \half \beta_0 \chi_0
\chi_0''\tilde{\gamma}_{s}'' \tilde{\gamma}_{ss}\right)\Big|_{M =
\tilde{\gamma}_s}.} 

The full NNLO running coupling correction eq.~\dgsss\ is given here
for the first time. In particular, the
term $\Delta \gamma_{sss}^{(0)}$, which depends 
only on $\chi_0$ and $\beta_0$, 
was already derived in ref.~\mf\ (eq.~(14) of that reference)
and  later confirmed in ref.~\ciafscheme\ [eq.~(4.16) of this 
reference, which gives ${\cal N}(N,t)$  eq.~\rcfactrc]. However, the terms
$\Delta \gamma_{sss}^{(1)}$, which  are  due to the next-to-leading order 
kernel $\chi_1$, and the last term on the right-hand side of eq.~\dgss,
due to the next-to-leading order running of the
coupling were never computed before.

\appendix{B}{Expansion of the GLAP anomalous dimensions}
The coefficients of the expansion eqs.~\gzexp-\gtexp\ of the leading,
next-to-leading and next-to-next-to-leading GLAP anomalous dimensions
in powers of $N$ are easily determined by recalling that $\gamma$ is
the large eigenvalue of the $2\times2$ anomalous dimension matrix, given
by
\eqn\largeev{
\gamma=\half\left[\gamma_{gg} + \gamma_{qq} +
\sqrt{(\gamma_{gg} - \gamma_{qq})^2
+4\gamma_{gq}\gamma_{qg}}\,\right],}
and using the expressions of $\gamma_{ij}$ given in
Refs.~\refs{\glap,\nlo,\nnlo}. In the \MS\ scheme we get

\eqn\gcoeffs{\eqalign{
g_{0,-1}&= \smallfrac{C_A}{\pi }\cr
g_{0,0}&=-\smallfrac{11 C_A}{12 \pi }+\left(-\smallfrac{1}{6 \pi }+\smallfrac{C_F}{3 \pi  C_A}\right) n_f\cr
g_{0,1}&=-\smallfrac{C_A \pi }{6}+\smallfrac{67 C_A}{36 \pi }-\smallfrac{11 C_F n_f}{36 \pi C_A}
+\left(-\smallfrac{C_F^2}{9 \pi C_A^3}+\smallfrac{C_F}{18 \pi C_A^2}\right) n_f^2\cr
g_{1,-1}&=\left(\smallfrac{13 C_F}{18 \pi ^2}-\smallfrac{23 C_A}{36 \pi ^2}\right) n_f\cr
g_{1,0}&=-\smallfrac{2 \zeta (3) C_A^2}{\pi ^2}+\smallfrac{1643 C_A^2}{216 \pi ^2}-\smallfrac{11 C_A^2}{36}+\left(\smallfrac{43
C_A}{54 \pi ^2}+\smallfrac{C_F}{18}-\smallfrac{547 C_F}{216 \pi ^2}+\smallfrac{C_F^2}{4 \pi ^2 C_A}\right) n_f\cr
&\qquad+\left(\smallfrac{13 C_F}{108 \pi ^2 C_A}-\smallfrac{13 C_F^2}{54 \pi ^2 C_A^2} \right) n_f^2\cr
g_{2,-2}&=\smallfrac{\zeta (3) C_A^3}{2 \pi ^3}+\smallfrac{11 C_A^3}{72 \pi }-\smallfrac{395 C_A^3}{108 \pi ^3}+\left(\smallfrac{
C_A^2}{36 \pi }-\smallfrac{71 C_A^2}{108 \pi ^3}-\smallfrac{C_F C_A}{18 \pi }+\smallfrac{71 C_F C_A}{54 \pi ^3}\right) n_f\cr
g_{2,-1}&=-\smallfrac{143 \zeta (3) C_A^3}{24 \pi ^3}-\smallfrac{29 \pi  C_A^3}{720}-\smallfrac{389 C_A^3}{432 \pi }+\smallfrac{73091
C_A^3}{2592 \pi ^3}+\left(-\smallfrac{11 \zeta (3) C_A^2}{12 \pi ^3}-\smallfrac{C_A^2}{9 \pi }\right.\cr
&\qquad\left.+\smallfrac{301 C_A^2}{81 \pi ^3}+\smallfrac{8 \zeta(3) C_F C_A}{3 \pi ^3}+\smallfrac{35 C_F C_A}{108 \pi }-\smallfrac{28853 C_F C_A}{2592 \pi ^3}
-\smallfrac{2 C_F^2 \zeta (3)}{3 \pi ^3}+\smallfrac{11 C_F^2}{12 \pi ^3}\right) n_f\cr
&\qquad\qquad+\left(\smallfrac{59 C_A}{648 \pi ^3}-\smallfrac{65 C_F}{324 \pi ^3}\right) n_f^2.}}

\appendix{C}{The ${\cal R}$ scheme change to next-to-next-to-leading order.}

The ${\cal R}$ scheme change~\refs{\ciafmsb,\CH,\ciafscheme} 
is related to the fact that the \MS\
anomalous dimension contains interference terms between collinear
poles and the $\beta$ function in $d$ dimensions. The factorization of
collinear singularities for a $d$--dimensional partonic cross section $\hat{\sigma}$ 
which depends on a single dimensionful variable $Q^2$ 
can be
written as
\eqn\dfact{
{\sigma}\left(\frac{Q^2}{\mu^2},\as(\mu^2),N,\varepsilon\right)= 
{\sigma^{(0)}}(Q^2,\alpha_0,N,\varepsilon) \exp 
\left[\int_{0}^{\as(\mu^2)}d \alpha
\frac{\gamma(\alpha,N,\varepsilon)}{\beta(\alpha,\varepsilon)}\right],
}
where  $\as(\mu^2)$ is the dimensionless
renormalized coupling, $\alpha_0$ is the bare coupling,
${\sigma^{(0)}}(Q^2,\alpha_0,N,\varepsilon)$ is the regularized
cross section and
${\sigma}(\frac{Q^2}{\mu^2},\as(\mu^2),N,\varepsilon)$ is free of
collinear singularities. The factorization scale is $\mu^2$ and
$\gamma(\as,N,\varepsilon)$ and $\beta(\as,\varepsilon)$ are
respectively the $d$--dimensional anomalous dimension and beta
function. The former is defined as the Mellin transform of the
$d$--dimensional Altarelli--Parisi splitting function. The latter is given by
\eqn\betad{\beta(\as,\varepsilon)=
\as\varepsilon+\beta(\as)}
in terms of the usual four--dimensional $\beta$ function
\eqn\betafour{\beta (\as)= -\beta_0
\as^2 (1+\as \beta_1 + \dots).}

The \MS\ anomalous dimension is the
residue of the simple pole in $\varepsilon$ in the integrand of the
exponential in eq.~\dfact, namely
\eqn\ddimcontr{\eqalign{ \gamma^{\ms}(\as,N)&={\hbox{Res}}_\varepsilon
\left[\frac{\as\gamma(\as,N,\varepsilon)}{\beta(\as,\varepsilon)}\right]\cr
&= \gamma(\as,N) - \frac{\beta(\as)}{\as} \dot\gamma(\as,N)+
\left(\frac{\beta(\as)}{\as}\right)^2\ddot\gamma(\as,N)+\dots,\cr
}} 
where the various coefficients are defined through the Taylor expansion
\eqn\gammafour{\gamma(\as,N,\varepsilon)\equiv\gamma(\as,N)
+\varepsilon\dot\gamma(\as,N)+\varepsilon^2\ddot\gamma(N)+\dots.
}
The \MS\ 
anomalous dimension thus receives two different classes of
contributions: pure collinear singularities and interference terms
between the $\varepsilon$-dependent anomalous dimension and the
poles arising from the expansion of the $d$-dimensional
$\beta$-function. In particular up to next-to-next-to leading
order we have
\eqn\fullmsb{\eqalign{
 \gamma_0^{\ms}&= \gamma_0 \,, \cr
 \gamma_1^{\ms}&= \gamma_1 + \beta_0 \dot{\gamma}_0  \,, \cr
 \gamma_2^{\ms}&= \gamma_2 + \beta_0 \beta_1 \dot{\gamma}_0 +\beta_0^2 \ddot{\gamma}_0+\beta_0 \dot{\gamma}_1 .}}

The $\cal R$ scheme change, which appears in the relation between \MS\
and \Q0\ schemes eq.~\qzmsb, takes us from the \MS\ scheme to the \MSS\ scheme
where the anomalous dimension is simply given by $\gamma(\as,N,0)$, i.e. 
$\gamma_0(N)$, $\gamma_1(N)$, etc. Thus in order to compute
$\gamma^{\mss}$ we have to subtract from the $\MS$ anomalous
dimension eq.~\fullmsb\ the contributions coming from the interference between
the $d$-dimensional kernel and the $\beta$ function, which are in turn
determined from the knowledge of the $d$--dimensional anomalous dimension 
$\gamma(\as,N,\varepsilon)$.

The $d$ dimensional leading order splitting functions have been 
known for a long time, at least for $x<1$~\peps:\eqn\ddim{\eqalign{
P_{qq}(x,\varepsilon) &= C_F \frac{1}{(1-x)^{\varepsilon}}
\left[\frac{1+x^2}{1-x}-\varepsilon (1-x)\right] 
                 +a_{qq}(x,\varepsilon) \delta(1-x) \,,  \cr
P_{qg}(x,\varepsilon) &= C_F \frac{1}{(1-x)^{\varepsilon}}\left[\frac{1+(1-x)^2}{x}-\varepsilon x \right] \,, \cr
P_{gq}(x,\varepsilon) &= T_R \frac{1}{(1-x)^{\varepsilon}}\left[1-2 x \frac{1-x}{1-\varepsilon}\right] \,,  \cr
P_{gg}(x,\varepsilon) &= 2 C_A \frac{1}{(1-x)^{\varepsilon}} \left[\frac{x}{1-x}+\frac{1-x}{x}+ x (1-x) \right] 
                 +a_{gg}(x,\varepsilon) \delta(1-x) \,.
}}
The end--point contribution $a_{qq}$ ($a_{gg}$)  can be extracted from
any process with collinear radiation from
incoming quarks, such as Drell-Yan, or gluons, such as Higgs production from
gluon fusion: the $O(\as)$ coefficient of the 
$\delta(1-x)$  provides a determination of the end--point term in
the splitting function after factoring a simple 
$\varepsilon$ pole and the Born cross section (and a factor of two
when there are two incoming partons). Using the known
NLO corrections for
Drell-Yan~\dynlo\ and Higgs~\hnlo\ production we get  
\eqn\deltacoeff{\eqalign{
a_{qq}(\varepsilon) &= C_F
\left[\frac{2}{\varepsilon}+\frac{3}{2}+ \varepsilon
\left(4-\frac{\pi^2}{3}\right) \right]+O(\varepsilon^2)\,,
\cr a_{gg}(\varepsilon)&= \frac{2C_A}{\varepsilon}+
\frac{11 C_A-4 n_f T_R}{6}- \varepsilon
\pi^2 +O(\varepsilon^2)\,. }}
The simple $\varepsilon$ pole cancels against that coming from the
expansion of $(1-x)^{-(1+\varepsilon)}=\frac{1}{\varepsilon}\delta(1-x)+\dots$
in the splitting functions $P_{qq}$ and $P_{gg}$, thereby providing a
check of the result.

Using the splitting functions~\ddim-\deltacoeff\ we can now determine the 
coefficients of the expansion in powers of $N$ of the
large
eigenvalue of the 
$\varepsilon$-dependent GLAP anomalous dimension matrix: we find 
\eqn\dotlo{\eqalign{
\dot{\gamma}_0(N)&=\frac{\dot{g}_{0,-2}}{N^2}+\frac{\dot{g}_{0,-1}}{N}+\dot{g}_{0,0}+O(N)\,,
\cr \ddot{\gamma}_0 (N)&=\frac{\ddot{g}_{0,-1}}{N}+O(N^0)\,
 }}
with
\eqn\dotcoeff{\eqalign{
\dot{g}_{0,-2} &= 0,\qquad
\dot{g}_{0,-1} = 0, \cr
\dot{g}_{0,0} &=  -\frac{67}{12 \pi}-\frac{7}{81}\frac{n_f}{\pi},\qquad
\ddot{g}_{0,-1} =  -\frac{\pi^2}{12} \,.
}}

The next-to-leading order $d$ dimensional splitting function is not
available, though in principle it could be extracted from
$d$--dimensional splitting amplitudes~\kos.  
However, the $\cal R$ scheme change has been determined long
ago~\ciafmsb\ to the level of the next-to-leading $N=0$ singularities,
i.e. for $\gamma_{ss}(\as/N)$ eq.~\gamexp. This corresponds to the
$O(\varepsilon)$ correction to the leading $N=0$ singularities,
because~\ciafscheme  
\eqn\rnllx{\eqalign{ \gamma_s^{\ms} &= \gamma_s \,,
\cr \gamma_{ss}^{\ms} &= \gamma_{ss}+\beta_0 \dot{\gamma}_s ,
\cr }} 
where we have used a notation similar to that of eq.~\gammafour\ but
for the expansion eq.~\gamexp\ of the anomalous dimension. 
Hence, if we let
\eqn\gammaparam{\dot{\gamma}_1 =
\frac{\dot{g}_{1,-3}}{N^3}+\frac{\dot{g}_{1,-2}}{N^2}+\frac{\dot{g}_{1,-1}}{N}+O(N^0)
\,. }
the coefficients $\dot{g}_{1,-3}$ and $\dot{g}_{1,-2}$ can be
extracted using eq.~\rnllx\ from the scheme change of ref.~\ciafmsb.

Equation~\rnllx\ implies that  $\gamma_s$ is left unaffected by the
scheme change, so
it follows from eq.~\fullmsb\ that 
$\dot{g}_{0,-i}=\dot{g}_{0,-1}=\dot{g}_{1,-3}=0$ thereby confirming the result
of eq.~\dotcoeff. Also,
the scheme change~\ciafmsb\ of $\gamma_{ss}$ starts at 
$O\left( \as \left( \frac{\as}{N} \right) ^3\right)$
\eqn\dotgammas{
\dot{\gamma}_s\left(\frac{\as}{N}\right)=2
\zeta(3)\left(\frac{\as}{N}\right)^3
+O\left(\left(\frac{\as}{N}\right)^4\right).
} 
Collecting everything we get
\eqn\nldotcoeff{ \dot{g}_{1,-3}=0;\quad\dot{g}_{1,-2}=0,}
while the sub--subleading coefficient $\dot{g}_{1,-1}$ remains undetermined:
it would require knowledge of the $O(\varepsilon)$ correction to the
simple $N$--pole contribution to $\gamma_1(\as,N,\varepsilon)$.

Summarizing, the \MSS\ anomalous dimension is given in terms of the
coefficients eq.~\gcoeffs\ of the expansion of the \MS\ anomalous
dimension and of the scheme change coefficients
eqs.~\dotcoeff,\nldotcoeff\ by 
\eqn\rscheme{\eqalign{ \gamma_0^{\mss} &=
\frac{g_{0,-1}}{N}+g_{00}+g_{0,1}N+ O(N^2)  \,, \cr \gamma_1^{\mss}
&= \frac{g_{1,-1}}{N} + \bar{g}_{1,0} + O(N) \,, \cr
\gamma_2^{\mss} &= \frac{g_{2,-2}}{N^2} + \frac{\bar{g}_{2,-1}}{N}
+ O(N^0) \,, }} 
where 
\eqn\rscheme{\eqalign{ \bar g_{1,0} & =
g_{1,0} - \beta_0 \dot{g}_{0,0} \,,\cr \bar g_{2,-1} & = g_{2,-1} -
\beta_0 \dot{g}_{1,-1}- \beta_0^2 \ddot{g}_{0,-1} \,.}}

\appendix{D}{The BFKL kernel in the $\Q0$ scheme}
In this appendix we give explicit expressions for our approximation to
the NNLO BFKL
kernel in the \Q0\ factorization scheme. 
The kernel for evolution of the unintegrated distribution, with the
argument of the strong coupling chosen as $\as(Q^2)$, and the
symmetric choice of kinematic variables eq.~\bfklkdef\ is given by
\eqn\finalqa{\eqalign{
\chi_0(M) & = \smallfrac{C_A}{\pi}\big(\smallfrac{1}{M}+ \smallfrac{1}{(1-M)}-1-
M(1-M)\big);}} 
\eqn\finalqb{\eqalign{
\chi_1(M) & = -\smallfrac{C_A^2}{2 \pi^2}
\big(\smallfrac{1}{M^3}+\smallfrac{1}{(1-M)^3}\big)+ 
\smallfrac{C_A}{\pi} \big(-\smallfrac{11 C_A}{12 \pi}-\smallfrac{n_f}{6 \pi}+
\smallfrac{C_F n_f}{3 \pi C_A}\big)
\big(\smallfrac{1}{M^2}+\smallfrac{1}{(1-M)^2}\big) \cr 
&-\smallfrac{C_A}{\pi}\beta_0 \smallfrac{1}{(1-M)^2} +\big(\smallfrac{13 C_F n_f}{18
\pi^2}-\smallfrac{23 C_A n_f}{36 \pi^2}\big)
\big(\smallfrac{1}{M}+\smallfrac{1}{1-M}\big) \cr 
& + \big(\smallfrac{C_A^2 \zeta(3)}{\pi^2}-\smallfrac{\zeta(3)
C_A^2}{2\pi^2}+\smallfrac{11 C_A^2}{72}-\smallfrac{395 C_A^2}{108 \pi^2}+ 
\smallfrac{C_A n_f}{36}-\smallfrac{71 C_A n_f}{108 \pi^2}-\smallfrac{C_F n_f}{18} \cr
&+\smallfrac{71 C_F n_f}{54 \pi^2}+\beta_0 \smallfrac{C_A}{\pi}\big) 
 + \smallfrac{C_A^2}{2 \pi^2}-\smallfrac{C_A}{\pi} \big(-\smallfrac{11 C_A}{12
\pi}-\smallfrac{n_f}{6 \pi}+ \smallfrac{C_F n_f}{3 \pi C_A}\big) \cr &- 
\big(\smallfrac{13 C_F n_f}{18 \pi^2}-\smallfrac{23 C_A n_f}{36 \pi^2}\big)
-\smallfrac{C_A}{\pi} \beta_0 M;}} 
\eqn\finalqc{\eqalign{
\chi_2(M) & = \smallfrac{C_A^3}{2 \pi^3}
\big(\smallfrac{1}{M^5}+\smallfrac{1}{(1-M)^5}\big)- 
               \smallfrac{C_A^2}{2 \pi^2}\big(-\smallfrac{11 C_A}{4
\pi}-\smallfrac{n_f}{2 \pi}+ \smallfrac{C_F n_f}{\pi C_A} +\beta_0 \big)
\big(\smallfrac{1}{M^4}+\smallfrac{1}{(1-M)^4}\big) \cr 
	       &+4\beta_0 \smallfrac{C_A^2}{\pi^2}\smallfrac{1}{(1-M)^4}
	       +\smallfrac{C_A}{\pi}\big[\big(-\smallfrac{11 C_A}{12
\pi}-\smallfrac{n_f}{6 \pi}+ \smallfrac{C_F n_f}{3 \pi C_A}\big)^2 \cr &+ 
	       \smallfrac{C_A}{\pi}\big(-\smallfrac{C_F^2 n_f^2}{9 C_A^3
\pi}+\smallfrac{C_F n_f^2}{18 C_A^2 \pi}-\smallfrac{11 C_F n_f}{36 C_A \pi} -
\smallfrac{C_A \pi}{6}+\smallfrac{67 C_A}{36 \pi}\big)-\big(\smallfrac{13 C_F
n_f}{18 \pi^2}-\smallfrac{23 C_A n_f}{36 \pi^2}\big)\big] \cr &\cdot 
\big(\smallfrac{1}{M^3}+\smallfrac{1}{(1-M)^3}\big)+2
\smallfrac{C_A}{\pi}\beta_0\big(\beta_0 
	       +\smallfrac{11 C_A}{6 \pi}+\smallfrac{n_f}{3 \pi}- \smallfrac{2 C_F
n_f}{3 \pi C_A}\big)\smallfrac{1}{(1-M)^3} \cr &+ 
	       \big[-\smallfrac{1}{2}\big(\smallfrac{C_A^3 \zeta(3)}{2
\pi^3}+\smallfrac{11 C_A^3}{72 \pi}-\smallfrac{395 C_A^3}{108 \pi^3}+ 
\smallfrac{C_A^2 n_f}{36 \pi}-\smallfrac{71 C_A^2 n_f}{108 \pi^3}-\smallfrac{C_A C_F
n_f }{18 \pi}+\smallfrac{71 C_A C_F n_f}{54 \pi^3} \big) \cr 
&+\big(-\smallfrac{11 C_A}{12 \pi}-\smallfrac{n_f}{6 \pi}+ \smallfrac{C_F n_f}{3 \pi
C_A}\big) 
\big(\smallfrac{13 C_F n_f}{18 \pi^2}-\smallfrac{23 C_A n_f}{36 \pi^2}\big)\cr
&+\smallfrac{C_A}{\pi}
\big(-\smallfrac{2 \zeta(3) C_A^2}{\pi^2}+\smallfrac{1643 C_A^2}{216
\pi^2}-\smallfrac{11 C_A^2}{36}+\smallfrac{43 C_A n_f}{54 \pi^2}+ 
\smallfrac{C_F n_f}{18} -\smallfrac{547 C_F n_f}{216 \pi^2}+\smallfrac{13 C_F
n_f^2}{108 \pi^2 C_A}\cr & +\smallfrac{C_F^2 n_f}{4 \pi^2 C_A}-\smallfrac{13
C_F^2 n_f^2}{54 \pi^2 C_A^2}+\beta_0 \big(\smallfrac{67}{12 \pi}+\smallfrac{7
n_f}{81 \pi}\big) \big)+\smallfrac{3 \zeta(3) C_A^3}{2 \pi^3} \big]
\big(\smallfrac{1}{M^2}+\smallfrac{1}{(1-M)^2}\big)\cr  & -\beta_0
\big(\smallfrac{C_A}{\pi}\beta_1+2\big(\smallfrac{13 C_F n_f}{18
\pi^2}-\smallfrac{23 C_A n_f}{36 \pi^2}\big)\big)\smallfrac{1}{(1-M)^2} \cr 
&+ \big[-\smallfrac{143 \zeta(3) C_A^3}{24 \pi^3}-\smallfrac{29 \pi
C_A^3}{720}-\smallfrac{389 C_A^3}{432 \pi}+\smallfrac{73091 C_A^3}{2592 \pi^3} 
-\smallfrac{11 C_A^2 \zeta(3) n_f}{12 \pi^3}-\smallfrac{C_A^2 n_f}{9 \pi} \cr
&+\smallfrac{301 C_A^2 n_f}{81 \pi^3}+\smallfrac{8 \zeta(3) C_A C_F n_f}{3
\pi^3} 
+\smallfrac{35 C_A C_F n_f}{108 \pi}+\smallfrac{59 C_A n_f^2}{648
\pi^3}-\smallfrac{28853 C_A C_F n_f}{2592 \pi^3} \cr &-\smallfrac{2 \zeta(3)
C_F^2 n_f}{3 \pi^3} 
-\smallfrac{65 C_F n_f^2}{324 \pi^3}+\smallfrac{11 C_F^2 n_f}{12 \pi^3}-\beta_0
\dot{g}_{1-1}+\beta_0^2 \smallfrac{\pi^2}{12}-2 \beta_0 \smallfrac{\zeta(3)
C_A^2}{\pi^2}\big] 
\big(\smallfrac{1}{M}+\smallfrac{1}{1-M}\big).
}} 

Substituting the numerical values of the Casimirs $C_A=3$ and $C_F=\smallfrac{4}{3}$ we get
\eqn\finalexplicita{\eqalign{
\chi_0(M) &=\smallfrac{3}{\pi }\big(\smallfrac{1}{M} 
+\smallfrac{1}{1-M}-1-M(1-M)\big), 
}} 
\eqn\finalexplicitb{\eqalign{
\chi_1(M) & =-\smallfrac{9}{2\pi ^2}\big(\smallfrac{1}{M^3}+\smallfrac{1}{(1-M)^3}\big)
-\big(\smallfrac{33}{4 \pi ^2}+\smallfrac{n_f}{18 \pi ^2}\big)\smallfrac{1}{M^2}
-\big(\smallfrac{33}{2 \pi ^2} 
-\smallfrac{4 n_f}{9\pi ^2}\big)\smallfrac{1}{(1-M)^2} 
-\smallfrac{103 n_f}{108\pi ^2}\big(\smallfrac{1}{M}+\smallfrac{1}{1-M}\big) \cr
&\qquad\qquad +\smallfrac{11}{8}+\smallfrac{n_f}{108}-\smallfrac{143}{12 \pi ^2}+\smallfrac{47 n_f}{162
\pi ^2}+\smallfrac{27 \zeta(3)}{2 \pi^2}+M \big(-\smallfrac{33}{4 \pi
^2}+\smallfrac{n_f}{2 \pi ^2}\big),}} 
\eqn\finalexplicitc{\eqalign{
\chi_2(M) &= \smallfrac{27}{2\pi ^3}\big(\smallfrac{1}{M^5}
+\smallfrac{1}{(1-M)^5}\big)+\big(\smallfrac{99}{4 \pi ^3}
+\smallfrac{n_f}{\pi ^3}\big) \smallfrac{1}{M^4} 
+\big(\smallfrac{495}{4 \pi ^3}-\smallfrac{5 n_f}{\pi ^3}\big) \smallfrac{1}{(1-M)^4} \cr
&+\big[\smallfrac{1167}{16 \pi ^3}+\smallfrac{35 n_f}{18 \pi
^3}+\smallfrac{n_f^2}{108 \pi ^3}-\smallfrac{9}{2 \pi }\big] \smallfrac{1}{M^3} 
+\big[\smallfrac{1893}{16 \pi^3}+\smallfrac{23 n_f}{9 \pi ^3}-\smallfrac{7 n_f^2}{36
\pi ^3}-\smallfrac{9}{2 \pi }\big] \smallfrac{1}{(1-M)^3} \cr 
&+\big[\smallfrac{1653}{16 \pi ^3}+\smallfrac{377 n_f}{432 \pi ^3}-\smallfrac{5
n_f^2}{648 \pi ^3}+\smallfrac{99}{16 \pi }+\smallfrac{5 n_f}{24 \pi }-\smallfrac{243
\zeta(3)}{4 \pi ^3}\big]\smallfrac{1}{M^2} \cr 
&+\big[\smallfrac{1653}{16 \pi ^3}-\smallfrac{5049}{8 (33-2 n_f) \pi
^3}+\smallfrac{881 n_f}{144 \pi ^3}+\smallfrac{933 n_f}{8 (33-2 n_f) \pi
^3}-\smallfrac{211 n_f^2}{648 \pi ^3} \cr 
&-\smallfrac{19 n_f^2}{4 (33-2 n_f) \pi ^3}+\smallfrac{99}{16 \pi }+\smallfrac{5
n_f}{24 \pi }-\smallfrac{243 \zeta(3)}{4 \pi ^3}\big]\smallfrac{1}{(1-M)^2} \cr
&+\big[\smallfrac{121}{192}-\smallfrac{11
n_f}{144}+\smallfrac{n_f^2}{432}+\smallfrac{73091}{96 \pi ^3}-\smallfrac{6125
n_f}{648 \pi ^3}+\smallfrac{11 n_f^2}{1944 \pi ^3}-\smallfrac{389}{16 \pi
}-\smallfrac{11 
\dot{g}_{1-1}}{4 \pi } \cr
&+\smallfrac{8 n_f}{27 \pi }+\smallfrac{\dot{g}_{1-1} n_f}{6 \pi } -\smallfrac{87 \pi
}{80}-\smallfrac{1683 \zeta(3)}{8 \pi ^3}+\smallfrac{457 n_f \zeta(3)}{108 \pi
^3}\big] 
\big(\smallfrac{1}{M}+\smallfrac{1}{1-M}\big).}} 

The expression of the NNLO kernel in DIS variables can be obtained by
inverting eqs.~\csnsexa-\csnsexb, with the result 
\eqn\finaldisnlo{\eqalign{
\chi_2^{DIS} &= \big(-\smallfrac{9}{2 \pi}+\smallfrac{1167}{16 \pi ^3}-\smallfrac{11
n_f}{12 \pi ^3}+\smallfrac{n_f^2}{108 \pi ^3}\big) \smallfrac{1}{M^3} 
+\big(\smallfrac{863}{16 \pi ^3}+\smallfrac{33}{4 \pi}-\smallfrac{54 \zeta (3)}{\pi
^3}+\smallfrac{235 n_f}{432 \pi ^3}\cr 
& +\smallfrac{2 n_f}{9 \pi}-\smallfrac{5 n_f^2}{648 \pi ^3}\big) \smallfrac{1}{M^2}+
\big(\smallfrac{121}{192}+\smallfrac{73091}{96 \pi ^3}-\smallfrac{87 \pi }{80} 
-\smallfrac{389}{16 \pi}-\smallfrac{11 \dot{g}_{1-1}}{4 \pi}-\smallfrac{1287 \zeta
(3)}{8 \pi ^3} \cr 
&-\smallfrac{11 n_f}{144}+\smallfrac{8 n_f}{27 \pi}-\smallfrac{6125 n_f}{648
\pi ^3}+\smallfrac{\dot{g}_{1-1} n_f}{6 \pi}+\smallfrac{133 n_f \zeta (3)}{108
\pi ^3}+\smallfrac{n_f^2}{432} 
+\smallfrac{11 n_f^2}{1944 \pi ^3}\big) \smallfrac{1}{M}\cr
&+\smallfrac{54}{(1-M)^5 \pi^3}+\big(\smallfrac{1683}{8 \pi^3}-\smallfrac{31 n_f}{4
\pi^3} \big) \smallfrac{1}{(1-M)^4} \cr 
&+\big(-\smallfrac{9}{2 \pi}+\smallfrac{1893}{16 \pi ^3}+\smallfrac{65 n_f}{12 \pi
^3}-\smallfrac{7 n_f^2}{36 \pi^3}\big) \smallfrac{1}{(1-M)^3}\cr 
&+\big(\smallfrac{33}{8 \pi}+\smallfrac{2137}{16 \pi^3}-\smallfrac{135 \zeta (3)}{2
\pi ^3}+\smallfrac{7 n_f}{36 \pi}+\smallfrac{3811 n_f}{432 \pi^3} 
-\smallfrac{211 n_f^2}{648 \pi^3}\big) \smallfrac{1}{(1-M)^2}\cr
&+\big(\smallfrac{121}{192}+\smallfrac{73091}{96 \pi^3}-\smallfrac{389}{16 \pi
}-\smallfrac{87 \pi }{80}-\smallfrac{11 \dot{g}_{1-1}}{4\pi}-\smallfrac{1287 \zeta
(3)}{8\pi ^3} 
-\smallfrac{11 n_f}{144}+\smallfrac{8 n_f}{27 \pi}\cr
&-\smallfrac{6125 n_f}{648
\pi^3}+\smallfrac{\dot{g}_{1-1} n_f}{6 \pi }
+\smallfrac{133 n_f \zeta (3)}{108 \pi ^3}+\smallfrac{11 n_f^2}{1944
\pi^3}+\smallfrac{n_f^2}{432}\big) \smallfrac{1}{1-M}.
}}

Finally, the kernel for the evolution of the
integrated parton density can be obtained from the unintegrated one
through eq.~\iuiexp. Using DIS kinematics the difference at NNLO is given by
\eqn\expliciint{\eqalign{
\chi_2^{i}(M)-\chi_2^{u}(M) & =
\big( -\smallfrac{55 n_f}{18 \pi ^3}+\smallfrac{5 n_f^2}{27 \pi ^3} \big)
\smallfrac{1}{M^3} + \smallfrac{3 (-33+2 n_f)}{2 \pi ^3}\smallfrac{1}{(1-M)^3} \cr 
& +\smallfrac{12393-4938 n_f+206 n_f^2}{648 \pi ^3}\smallfrac{1}{M^2}+
\smallfrac{-15147+1182 n_f-16 n_f^2}{108\pi ^3}\smallfrac{1}{(1-M)^2} \cr 
& +\smallfrac{1}{3888 \pi ^3} \big[-703890+37974 n_f+284 n_f^2+29403 \pi ^2 \cr
&-1584 n_f \pi ^2-12 n_f^2 \pi ^2+96228 \zeta(3)-5832 n_f
\zeta(3)\big] \smallfrac{1}{M} \cr &-\smallfrac{24543+705 n_f-109 n_f^2}{324\pi
^3} \smallfrac{1}{1-M} .
}}  

\footatend\vfill\supereject\immediate\closeout\rfile\writestoppt
\baselineskip=14pt\centerline{{\bf References}}\bigskip{\frenchspacing%
\parindent=20pt\escapechar=` \input refs.tmp\vfill\eject}\nonfrenchspacing
\vfill\eject
\bye